\documentclass[prd,aps,a4,onecolumn,superscriptaddress,preprintnumbers,nofootinbib]{revtex4-1}

\usepackage{pslatex}
\usepackage[pdftex]{graphicx}
\usepackage{floatrow}
\usepackage{psfrag}
\usepackage{epsfig}
\usepackage{color}
\usepackage{cancel}
\usepackage{slashed}
\usepackage{amssymb}
\usepackage{amsmath}
\usepackage{hyperref}
\usepackage{enumerate}
\usepackage{multirow}
\usepackage{ulem}
\usepackage{float}
\usepackage{comment}
\usepackage{diagbox}
\usepackage{amsmath}
\usepackage{soul, xcolor}
\usepackage{makecell, parskip, booktabs}

\def\boxit#1{\vbox{\hrule\hbox{\vrule\kern6pt
			\vbox{\kern6pt#1\kern6pt}\kern6pt\vrule}\hrule}}

\def\bse{\begin{eqnarray*}}
	\def\ese{\end{eqnarray*}}
\def\be{\begin{eqnarray}}
	\def\ee{\end{eqnarray}}
\def\bq{\begin{equation}}
	\def\eq{\end{equation}}
\def\bse{\begin{eqnarray*}}
	\def\ese{\end{eqnarray*}}
\def\pr{\hbox{pr}}

\def\D{\mathcal{D}}
\def\M{\mathcal{M}}
\def\R{\mathcal{R}}

\def\T{\mathcal{T}}

\def\wh{\widehat}
\def\log{\hbox{log}}
\def\ln{\hbox{ln}}
\allowdisplaybreaks
\begin{document}
\setstcolor{red}

\title{Prospects for measuring time variation of astrophysical neutrino sources at dark matter detectors}

\author{Yi Zhuang}
\affiliation{Department of Physics and Astronomy, Mitchell Institute
  for Fundamental Physics and Astronomy, Texas A\&M University,
  College Station, Texas 77843, USA}
\author{Louis E. Strigari}%
\affiliation{Department of Physics and Astronomy, Mitchell Institute
  for Fundamental Physics and Astronomy, Texas A\&M University,
  College Station, Texas 77843, USA}

\author{Lei Jin}%
\affiliation{Department of Mathematics and Statistics, Texas A\&M University-Corpus Christi,  Corpus Christi,  TX 78412, USA}

\author{Samiran Sinha}%
\affiliation{Department of Statistics, Texas A\&M University,
  College Station, Texas 77843, USA}
  
\date{\today}
\begin{abstract}
We study the prospects for measuring the time variation of solar and atmospheric neutrino fluxes at future large-scale Xenon and Argon dark matter detectors. For solar neutrinos, a yearly time variation arises from the eccentricity of the Earth's orbit, and, for charged current interactions, from a smaller energy-dependent day-night variation to due flavor regeneration as neutrinos travel through the Earth. For a 100-ton Xenon detector running for 10 years with a Xenon-136 fraction of $\lesssim 0.1\%$, in the electron recoil channel a time-variation amplitude of about 0.8\% is detectable with a power of 90\% and the level of significance of 10\%. This is sufficient to detect time variation due to eccentricity, which has amplitude of $\sim 3\%$. In the nuclear recoil channel, the detectable amplitude is about 10\% under current detector resolution and efficiency conditions, and this generally reduces to about 1\% for improved detector resolution and efficiency, the latter of which is sufficient to detect time variation due to eccentricity. Our analysis assumes both known and unknown periods. We provide scalings to determine the sensitivity to an arbitrary time-varying amplitude as a function of detector parameters. Identifying the time variation of the neutrino fluxes will be important for distinguishing neutrinos from dark matter signals and other detector-related backgrounds, and extracting properties of neutrinos that can be uniquely studied in dark matter experiments. 
\end{abstract}
\maketitle

\section{Introduction \label{sec:introduction}}

\par Over the past several decades, direct dark matter detection experiments have made tremendous progress in constraining weak-scale particle dark matter~\cite{XENON:2018voc,LUX-ZEPLIN:2022qhg}. Future larger-scale detectors will be sensitive to not only particle dark matter, but also astrophysical neutrinos and various other rare-event phenomenology~\cite{Aalbers:2022dzr}. The most prominent of the neutrino signals are from the Sun, the atmosphere, and the diffuse supernova neutrino background (DSNB)~\cite{Billard:2013qya}. Understanding these signals has important implications for the future of particle dark matter searches, but also for understanding the nature of the sources and the properties of neutrinos~\cite{Dutta:2019oaj}. 

\par Various methods have been proposed to distinguish neutrinos and a possible dark matter signal. These include exploiting the energy distribution of nuclear recoils between neutrinos and dark matter~\cite{Dent:2016iht,Dent:2016wor}, the differences in arrival directions~\cite{OHare:2015utx}, and the differences in the periodicities of the signal~\cite{2015Davis,2021Sassi}. New physics in the neutrino sector may also change the nature of the predicted neutrino signal~\cite{Cerdeno:2016sfi,AristizabalSierra:2019ykk,AristizabalSierra:2021kht}, and provide a method to discriminate from dark matter. 

\par Here we examine the time variations of the neutrino signals in more detail and study the prospects for measuring these time variations. For solar neutrinos, the time variation of the flux is due to the eccentricity of the Earth's orbit and the day-night effect. The former is independent of the neutrino flavor, while the latter, which results from neutrino interactions with the matter as they pass through the Earth, is flavor dependent. Both effects are present in a dark matter experiment through the nuclear recoil and electron recoil channels. Beyond solar neutrinos, there may be detectable time variation of other components of the astrophysical neutrino flux. Due to the solar cycle, there is a time variation of the atmospheric neutrino flux which is $\sim 10-30\%$~\cite{Zhuang:2021rsg} depending on the detector location. We present the first estimates of the detectability of the time variation of the atmospheric neutrino flux given realistic future detector configurations. 

\par Identifying the time variation for the different neutrino sources is important for properly extracting the signal and distinguishing it from dark matter~\cite{Billard:2013qya,OHare:2021utq}. In addition, it is important to characterize these signals to constrain neutrino properties. Previous studies of neutrinos in dark matter experiments have only considered the time variation of solar neutrinos as being due to the eccentricity of the Earth's orbit, and for idealistic values of the nuclear recoil threshold~\cite{2015Davis}. In addition to examining more realistic and updated detector configurations, we consider the prospects for measuring time variation of solar neutrinos using the neutrino-electron elastic scattering channel for the first time. 

\par Solar neutrino experiments have previously searched for time variations in their signals, including experiments that have successfully established time variation due to the eccentricity of the Earth's orbit~\cite{Borexino:2022khe,2006:Kamiokande} and the day-night effect~\cite{Super-Kamiokande:2016yck}. These experiments each used a range of statistical technqiues to identify the time variable signal. As part of our analysis, we rigorously compare the different statistical methodologies for extracting the time-varying signal. We present results for the sensitivity of given experiments to time-varying amplitudes, and quantify the prospects for signal extraction as a function of experimental sensitivity and background levels. 

\par This paper is organized as follows. In Section~\ref{sec:SandB}, we briefly describe the signals and the models for the detector efficiency. In Section~\ref{sec: physical_Ps} we describe the periodic signals used in this work, and also we summarize the previous experiments that have searched for neutrino periodicity. Next, in Section~\ref{sec:statistical}, we review the statistical methodologies used in our analysis. In Section~\ref{sec:data_analysis}, we describe the simulation strategy, compare statistical methods and introduce the signal-to-noise ratio as a convenient estimation tool. Then in Section~\ref{sec:results}, we present our resulting projections, and in Section~\ref{sec:conclusions}, the discussion and conclusions. 


\section{Event rates at Different Detectors \label{sec:SandB}}

\subsection{Theoretical calculation}
\par
\begin{figure*}[!htbp]
\includegraphics[width = 0.95\textwidth ]{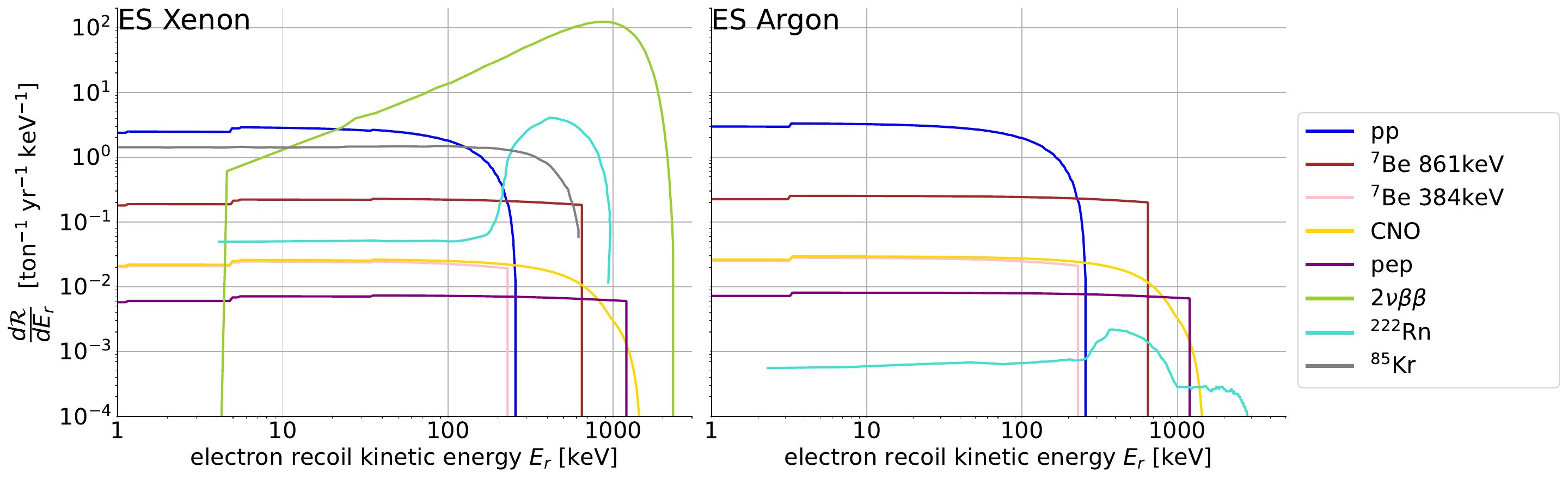}
\caption{Neutrino-electron elastic scattering (ES) spectra for Xenon (left) and Argon (right) for solar and experimental background components. For ES Argon, we consider $^{222}$Rn as the background~\citep{2016Franco}.
}
\label{fig: pdf_ES_XeAr}
\end{figure*}

\begin{figure*}[!htbp]
\includegraphics[width = 0.95\textwidth ]{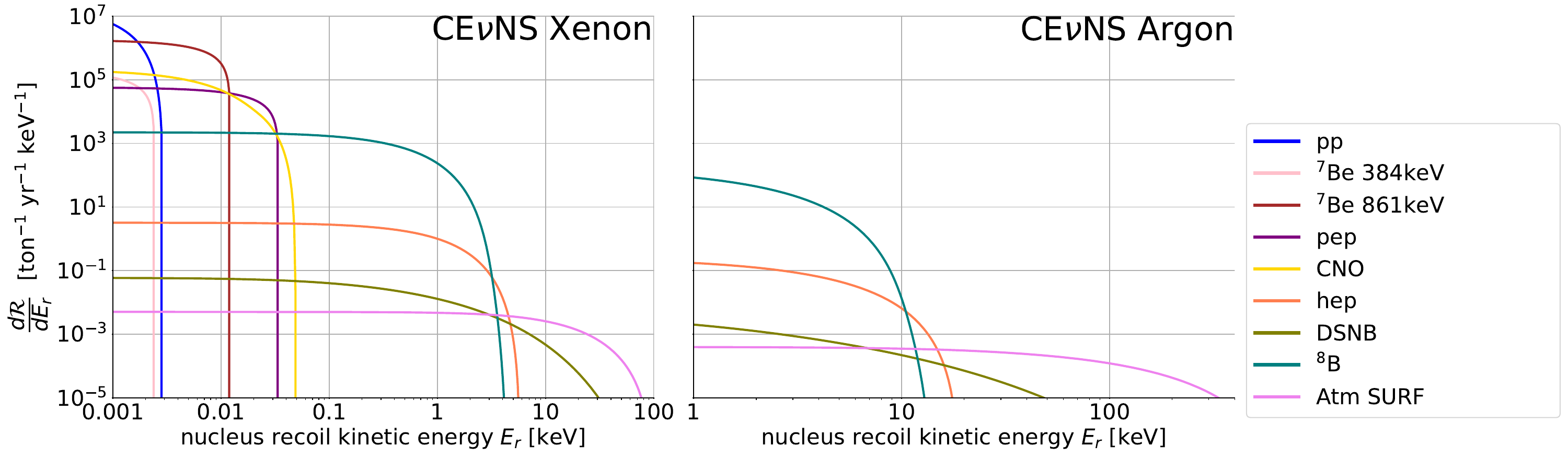}
\caption{ Neutrino-nucleus scattering (CE$\nu$NS) spectra for Xenon (left) and Argon (right). Shown are the components of the solar, atmospheric, and DSNB spectra. The atmospheric spectra are shown for the SURF detector location.
}
\label{fig: pdf_CEVENS_XeAr}
\end{figure*}

  Figures~\ref{fig: pdf_ES_XeAr} and ~\ref{fig: pdf_CEVENS_XeAr} show the electron and nuclear recoil spectra for the solar, atmospheric, and DSNB spectra for Xenon and Argon targets. The nuclear recoil spectrum uses the neutral current coherent elastic neutrino-nucleus scattering (CE$\nu$NS) channel, and the electron recoil channel uses charged current neutrino-electron elastic scattering (ES). We refer to previous literature for details of these calculations~\cite{Zhuang:2023dzd}. Here we simply highlight the components of the spectrum as a function of recoil energy to get a sense of the recoil threshold required to detect each component. For Solar neutrinos, we used the high metallicity model for the normalization~\cite{2013Haxton}. We also show the appropriate experimental background components~\cite{2019Jayden}. 

\par In addition to the energy dependence shown in Figure~\ref{fig: pdf_ES_XeAr} and~\ref{fig: pdf_CEVENS_XeAr}, for several of the neutrino components there is a time variation to the flux. For solar neutrinos, since the time variation due to the eccentricity of the Earth's orbit affects all flavors, it will be present in both the CE$\nu$NS and the ES channels. For charged current detection channels, the day-night effect due to oscillations is present in the ES channel. For atmospheric neutrinos, the time variation from the solar modulation of the atmospheric neutrino flux~\cite{Zhuang:2021rsg} affects all flavors and is detectable through CE$\nu$NS. Of the spectra shown in Figure~\ref{fig: pdf_CEVENS_XeAr}, the only essentially steady-state component is the DSNB, which has a cosmological origin~\cite{Suliga:2021hek}. 
In the section below we describe our parameterization of each of these time-varying components in more detail. 

\subsection{Adding Resolution and Detector Efficiency }
\par To estimate the detector efficiency and resolution in Xenon, we use the Noble Element Simulation Technique (NEST)~\citep{2011NEST} code. Neutrinos (or dark matter particles) interact with the gas in the detector, producing a scintillation signal, S1, and ionization electrons, which then drift along the electric field to produce a scintillation signal, S2. The NEST code simulates the detection of events in the space of S1 and S2. For the NEST configurations, we compare several possible options. Similar to previous studies, we consider the all enhanced parameters ``Xe100t-5"\footnote{\url{https://zenodo.org/records/3653516}``Detector\_Xe100t\_5.hh" }~\citep{2021Jayden}. We also use the specific experimental files LZ\footnote{\url{https://github.com/NESTCollaboration/nest/blob/master/include/Detectors/LZ_SR1.hh}} and G3\footnote{\url{https://github.com/NESTCollaboration/nest/blob/master/include/Detectors/Detector_G3.hh}}. For all cases, we do not select a specific region of interest (ROI), so the signal encompass all available space for electron and nuclear recoils. 

\par With this setup, we estimate the detector efficiency as a function of the true nuclear or electron recoil energy. To obtain the efficiency, we simulate $10^7$ random recoil energies uniformly over the range $[0-100]$ keV within the detector. After processing through NEST, each simulated energy corresponds to a specific S1 and S2. We bin the detected events in recoil energy space, count the number of events with both valid S1/S2~\citep{XENON:2020rca}, and divide the valid counts by the total number of events in each energy bin.

\begin{figure}[!htbp]
\includegraphics[width = 0.95\textwidth ]{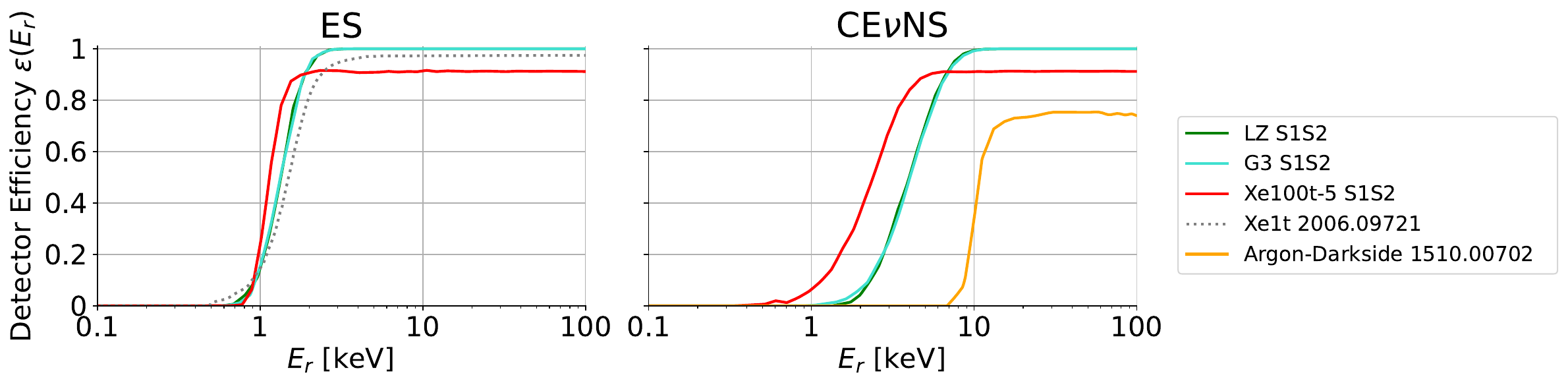}
\caption{Different detector efficiency models from NEST simulations (``LZ", ``G3", ``Xe100t-5") and current experiments (``Argon-DarkSide", ``Xe1T"). {\it Left panel}: Solid curves are ``LZ", ``G3" and ``Xe100t-5" for Xenon elastic scattering (ES). The dotted curve is the efficiency from Xenon1T~\citep{XENON:2020rca}. {\it Right panel}: Solid curves are ``LZ", ``G3" and ``Xe100t-5" for the Xenon nuclear recoil CE$\nu$NS channel. The orange solid curve is the efficiency from DarkSide-50~\citep{2016PhRvD..93h1101A} for nuclear recoils in Argon. 
} 
\label{fig:detection_accep}
\end{figure}

\par The results for the ES and CE$\nu$ES efficiencies are shown in  Figure~\ref{fig:detection_accep}. For comparison, we show the ES efficiency used by Xenon1T~\citep{XENON:2020rca}, which is similar to our ``Xe100t-5", LZ and G3 curves, with the differences between them likely explained by using different detector parameters, including the ROI and the recoil energies simulated. For Xenon, the electron recoil efficiency drops to zero rapidly below $\sim 1$ keV, and is nearly 100\% above this energy. 

\par To obtain the event rate modified by resolution and detector efficiencies, we integrate from a threshold energy, $E_{thrd}$, to an endpoint energy, $E_{end}$, and obtain $\R = \int_{E_{thrd}}^{E_{end}} (d\R/dE_{r}) dE_{r}$, where
\begin{equation}
    \frac{d\R}{dE_{r}} = \epsilon (E_{r}) \int dE_{r}^{'} \frac{d\R(E_{r}^{'})}{dE_{r}^{'}}
     \frac{1}{\sqrt{2\pi \sigma^{2} (E_{r}^{'})} } e^{-\frac{1}{2}\frac{(E_{r} - E_{r}^{'})^2}{ \sigma ^{2}  (E_{r}^{'})}}
    \nonumber
\end{equation}
where $E_{r}^{'}$ is the true recoil energy, $E_{r}$ is the detected recoil energy, $\sigma(E_{r}^{'})$ is the resolution at the true recoil energy, and $\epsilon (E_{r})$ is the detector efficiency at the detected recoil energy. 

\par Different resolution and efficiency models for different nuclear targets that we use for our analysis are summarized in Tables~\ref{tab:component_ES} and~\ref{tab:component_CEVENS}. For Xenon our resolution and efficiency models rely on NEST simulations as described above. For our Argon ES resolution and efficiency model, we choose $E_{thrd} = 100$ keV in order to determine how our results are affected by choosing a threshold characteristic of current experiments.  

\begin{table}[!htbp]
\caption{Target, resolution model, detector efficiency model, recoil energy threshold, recoil energy endpoint, components experiencing time variation, and components that have constant rate constant in time in the ES channel. $f_{2\nu\beta\beta}$ is the fraction of the remaining $2\nu\beta\beta$ background after depletion of $^{136}$Xe.  }
\begin{tabular}{c|c|c|c|c|c|c|c}
\hline
channel & target & resolution  model  & efficiency model & $E_{thrd}$ & $E_{end}$ & signal components & background components\\
 &  & $\sigma(E_{r})$ & $\epsilon(E_{r})$ & [keV]& [keV] & [ton$^{-1}$ yr$^{-1}$] & [ton$^{-1}$ yr$^{-1}$]\\
\hline
\hline 
\multirow{4}{*}{ES} & \multirow{2}{*}{Xe} & ideal & ideal  & 1 & 650 & \multirow{4}{*}{\makecell{pp, \\ $^{7}$Be 861, $^{7}$Be 384, \\pep, \\CNO}} &  \multirow{3}{*}{$f_{2\nu\beta\beta} \times 2\nu\beta\beta$, $^{85}Kr$, $^{222}Rn$}\\
\cline{3-6}

&  & $\left(0.31\sqrt{\frac{E_{r}}{\textrm{keV}}} + 0.0035\frac{E_{r}}{\textrm{keV}}\right) \textrm{keV}$~\cite{XENON:2020rca}  
& Xe100t-5 & 0 &  790  & & \\
\cline{2-6}
\cline{8-8}
&  \multirow{2}{*}{Ar}  & ideal & ideal & 1 & 3400  && \multirow{2}{*}{$^{222}Rn$} \\
\cline{3-6}
&   & $0.1E_{r}$ & / & 100 & 3400 &&\\
\hline
\end{tabular}
\label{tab:component_ES}
\end{table}

\begin{table}[!htbp]
\caption{Same as Table~\ref{tab:component_ES} except for the CE$\nu$NS channel. }
\begin{tabular}{c|c|c|c|c|c|c|c}
\hline
channel & target & resolution  model  & efficiency model & $E_{thrd}$ & $E_{end}$ & signal components & background components\\
 &  & $\sigma(E_{r})$ & $\epsilon(E_{r})$ & [keV]& [keV] & [ton$^{-1}$ yr$^{-1}$] & [ton$^{-1}$ yr$^{-1}$]\\
\hline
\hline 
\multirow{4}{*}{CE$\nu$NS} & \multirow{2}{*}{Xe} & ideal & ideal  & 1 & 4 & \multirow{4}{*}{$^{8}$B} & \multirow{4}{*}{/}\\
\cline{3-6}
&  & \multirow{3}{*}{$\left(0.31\sqrt{\frac{E_{r}}{\textrm{keV}}} + 0.0035\frac{E_{r}}{\textrm{keV}}\right) \textrm{keV}$~\cite{XENON:2020rca} 
} & Xe100t-5 & \multirow{3}{*}{0} &  \multirow{3}{*}{9}  & & \\
&  & & LZ &&& & \\
&  & & G3 &&& & \\
\cline{2-6}
&  \multirow{2}{*}{Ar}  & ideal & ideal & 1 & 13 &&\\
\cline{3-6}
&   & $0.1E_{r}$ & Darkside & 0 & 29 &&\\
\hline
\end{tabular}

\label{tab:component_CEVENS}
\end{table}

\begin{table}[!htbp]
\caption{Summary of previous experimental results searching for time variation of solar neutrinos. The columns are 1) Experiment, 2) signal of interest experiment searched for, 3) experimental exposure, 4) time binning used in the analysis, 5) scanned frequency range, 6) results of the analysis, along with specific comments where appropriate. Boxes are left blank in the experimental exposure and time binning columns for Homestake because that information was not provided. Boxes are left blank in the scanned frequency range column because in these cases the signal is fit for given the known eccentricity, and a scan of frequency is not required. 
}
\begin{tabular}{|c|c|c|c|c|c|}
\hline
Experiment  & Signals of interest  & $\T$ & $\Delta$ t  & \makecell{scanned frequency range \\ $[f_{min}, f_{max}]$} &  \makecell{results} \\

\hline
\hline 
\multirow{2}{*}{\makecell{Homestake \\~\cite{Sturrock:Homestake}}} & \makecell{quasi-biennial periodicity $\frac{1}{2.2}$ yr$^{-1}$, 
\\ Rieger periodicity $\frac{1}{157}$ day$^{-1}$
} & & &  [$0$, 20] yr$^{-1}$ &   no evidence  \\
\cline{2-6}
&  solar rotation frequency & & &  [12.4, 13.1] yr$^{-1}$ &  \makecell{peak at 12.88 yr$^{-1}$\\ significant at the 3\% level} \\
\hline
\hline 
\multirow{4}{*}{Sudbury}& blind search &\multirow{4}{*}{\makecell{572.2 calendar days (D$_2$O)\\  762.7 calendar days (salt)\\~\cite{2005:Sudbury,2007:SNO}}} & \multirow{6}{*}{\makecell{1 day}}& \makecell{[$\frac{1}{10}$, 365.25] yr$^{-1}$}& \makecell{peak at \\ 0.296 day$^{-1}$ (D$_2$O),\\ 0.971 day$^{-1}$ (salt),\\ 0.417 day$^{-1}$ (combined)}  \\

\cline{5-6}

&&&& \makecell{ 
[$\frac{1}{10}$, 182.625] yr$^{-1}$ }& \makecell{peak at \\  0.408 day$^{-1}$ (D$_2$O),\\  0.429 day$^{-1}$ (salt),\\ 0.413 day$^{-1}$ (combined)}  \\
\cline{2-2}\cline{5-6}
& \makecell{ 7\% amplitude 9.43 yr$^{-1}$ \\ Kamioka result} &&&[9.33, 9.53] yr$^{-1}$ &  \makecell{best-fit amplitude 1.3\% \\ (combined)}\\
\cline{2-2}\cline{5-6}
&eccentricity 1 yr$^{-1}$ &&&&\makecell{best-fit
 eccentricity\\ $\epsilon$=0.0143}\\

\hline
\hline 
\multirow{10}{*}{Borexino}&  \makecell{diurnal modulation 1 day$^{-1}$ }& 
\makecell{ 740.88 live days \cite{Borexino:2011bhn} \\ Phase-I}&& \/ & \makecell{day-night
\\ asymmetry = 0.001} \\
\cline{2-6}
&\multirow{3}{*}{eccentricity 1 yr$^{-1}$} &\multirow{3}{*}{\makecell{ 1456 astro days~\cite{2017:Borexino} \\ Phase-II }}&    30.43 days && \makecell{ best-fit
 eccentricity\\ $\epsilon$ =
0.0174} \\
\cline{4-6}
&&&\makecell{ 1 day\\ 7 days} && \makecell{peak at 1 yr$^{-1}$} \\
\cline{2-6}
&eccentricity 1 yr$^{-1}$ & \makecell{$\sim$ 10 yrs ~\cite{Borexino:2022khe} \\  Phase-II+III}&  \makecell{ 30 days }&& \makecell{ best-fit
 eccentricity\\ $\epsilon$ =
 0.0184}\\
 \cline{2-2} \cline{4-6}
&\makecell{Carrington rotation 13.4 yr$^{-1}$, \\ diurnal modulation 1 day$^{-1}$}& &  \makecell{ 8 hrs }& [1, 547] yr$^{-1}$& \makecell{day-night \\
 asymmetry = 0.003}\\
\hline
\hline 
\multirow{5}{*}{Kamioka}& blind search& \makecell{1871 elapsed days\\ SK-1 \\~\cite{2003PhRvD..68i2002Y,2003Sturrock,2006:Kamiokande,2005Sturrock,2006:SK1} } & \multirow{2}{*}{10 days} & \makecell{[0.0002, 0.0987] day$^{-1}$ \\ ~\cite{2003PhRvD..68i2002Y}}& \makecell{peak at 0.0726 day$^{-1}$}\\
\cline{5-6}
&& && [0, 50] yr$^{-1}$ ~\cite{2006:Kamiokande}& peak at 9.42 yr$^{-1}$\\
\cline{5-6}
&& && [0, 100] yr$^{-1}$ ~\cite{2003Sturrock}& \makecell{peak at 26.57 yr$^{-1}$ }\\
&& && [0, 40] yr$^{-1}$ ~\cite{2003Sturrock}& \makecell{ alias peak at  9.42 yr$^{-1}$}\\

\cline{4-6}
&& & \multirow{2}{*}{5 days} & \makecell{[0.0002, 0.19187]  day$^{-1}$ \\~\cite{2003PhRvD..68i2002Y}}& \makecell{peak at 0.1197 day$^{-1}$\\0.0726 day$^{-1}$ removed}\\
\cline{5-6}
&& & & [0, 50] yr$^{-1}$~\cite{2006:Kamiokande,2005Sturrock} & \makecell{alias peak of 9.42 yr$^{-1}$ \\ at 26.52yr$^{-1}$ \\possible solar rotation}\\
\cline{2-2}\cline{4-6}
& eccentricity 1 yr$^{-1}$ ~\cite{2006:SK1}&& 1.5 months & & \makecell{best-fit
 eccentricity\\ $\epsilon$=0.021}\\
\cline{2-2}\cline{4-6}
& \makecell{Day-Night difference ~\cite{2006:SK1}}&& & &\makecell{day-night
\\ asymmetry = -0.021, -0.017} \\

\cline{2-6}
& blind search &
\makecell{  5803 live days~\cite{Kamioka:2023}\\ SK-I,II, III, IV}&  \makecell{5 days} &\makecell{ [4e$^{-4}$, 0.2] day$^{-1}$ } & \makecell{peak at  0.126 day$^{-1}$} \\
\cline{5-6}
&&&  &\makecell{ [10$^{-6}$, 0.2] day$^{-1}$  } & \makecell{peak at   0.143 day$^{-1}$} \\
\cline{2-2}\cline{4-6}
&\makecell{eccentricity 1 yr$^{-1}$ }& &  $\frac{365.25}{12}$ days & & \makecell{best-fit
 eccentricity\\ $\epsilon$=0.0153} \\

\hline
\hline 
\multirow{2}{*}{ \makecell{IceCube \\ \cite{2019IceCube}}} & \multirow{2}{*}{ \makecell{temperature dependent variation \\ of atmospheric neutrino flux}}  & \multirow{2}{*}{May 2012  - May 2017 } & \multirow{2}{*}{ \makecell{ 1 day \\ 30 days} } &   & \multirow{2}{*}{ \makecell{high statistics at 140 day$^{-1}$ \\ in the South region} }\\
&&&&&\\
\hline
\end{tabular}
\label{tab:experiments}
\end{table}

\section{Physical Periodic Signals of Interest}\label{sec: physical_Ps}
\subsection{Solar yearly modulation} 
\par The slight eccentricity, $\epsilon$, to the Earth's orbit around the Sun induces a yearly variation in the neutrino flux. The time dependence of the flux is parameterized as~\citep{2013Borexino}
\begin{equation}
    \phi(t) = \frac{R_{\odot}}{4\pi r^{2}(t)} \approx \frac{R_{\odot}}{4\pi r_{0}^{2}} \left[1 + 2\epsilon \cos\left( \frac{2\pi t}{P}\right) \right],
    \label{eq:eccentricity}  
\end{equation}
where  $P$ denotes the periodicity measured in  years, 
$r(t)$ is the distance between the Earth and Sun at time $t$  measured in years (i.e., $0\leq t\leq 1$), $r_0 = 1$ AU, and $R_{\odot}$ is the neutrino production rate in the Sun per unit time.
The second term, $2\epsilon \cos( 2\pi t/P)$, in Equation~\ref{eq:eccentricity} is responsible for the small time variation. The amplitude of the flux variation is approximately 0.03342, and has been measured by several experiments, including recently by Borexino~\cite{Borexino:2022khe}. For standard assumptions for the dark matter velocity distribution, this time variation is out of phase with the standard modulation signal predicted from dark matter and has been proposed as a means to separate neutrinos from dark matter signals~\cite{2015Davis}.

\subsection{Solar daily modulation} 
\par A second time variation in the solar neutrino flux is the day night effect, which is due to the regeneration of the electron neutrino flux from matter effects as electron neutrinos pass through the Earth. The day-night asymmetry is detectable for experiments sensitive to the electron neutrino flux through charged current interactions. Defining $N_{D}$ and $N_{N}$ as the number of day and night events, the day-night asymmetry,  $A_{DN}$, is parametrized as~\cite{Borexino:2022khe} 
\begin{equation}
    A_{DN} = \frac{N_{D} - N_{N}}{0.5 (N_{D} + N_{N})}=\frac{2A_{d, DN}}{\sqrt{2}R_{\odot}},
    \nonumber
\end{equation}
where $A_{d, DN}$ is the amplitude of the daily modulation. Super-Kamiokande has now established the day-night effect for high energy ${}^8$B neutrinos~\cite{Super-Kamiokande:2016yck}  with a measured amplitude of $\lesssim 3\%$ that is consistent with the LMA-MSW solution. 

\par At lower energies where solar neutrinos transition are due to vacuum oscillations, the LMA-MSW solution predicts that the day-night modulation is much smaller, $\lesssim 0.1\%$. The best limits on the day-night modulation at these energies come from Borexino measurements of the ${}^7$Be component~\cite{Borexino:2022khe}. Borexino is sensitive to electron recoils $\gtrsim 150$ keV, so is not sensitive to the modulation for the lower energy pp component. Since Xenon and Argon-based dark matter experiments are sensitive to lower threshold~, $\sim$ few keV electron recoils, they will extend limits on solar neutrino time variation and day-night effects to lower energies than have been previously studied. In our analysis, we simply compare the limits on daily modulation from dark matter detectors to the limits on the modulation obtained from Borexino. 

\subsection{Atmospheric neutrinos} \label{subsec:atm}

\par We also study the prospects for identifying the time variation of the atmospheric neutrino flux. Atmospheric neutrinos are detected through the CE$\nu$NS channel, with a modulation amplitude that depends on the detector location. We model the atmospheric time variation with a sinusoidal function in a similar manner to Equation~\ref{eq:eccentricity}. In this case, the period is given by the $11$-year period of the Solar cycle, and the amplitudes are given by the maximum difference between the Solar minimum and maximum fluxes calculated in Ref.~\cite{Zhuang:2021rsg}. The flux differs for different detector locations, which we consider in our analysis below. For example for the detector locations China Jinping Underground Laboratory (CJPL), Kamioka, Laboratori Nazionali del Gran Sasso (LNGS), the Sanford Underground Research Facility (SURF), and SNOlab, the maximum amplitudes for time variation are $A_{atm} = [0.0459,0.0382,0.0461,0.1475,0.1327]$, respectively, where $A_{atm}$ and $\omega$ are defined through 
\begin{eqnarray*}
    \phi_{atm}(t) &=& \bar \phi \, A_{atm} \, \sin (\omega t ),\\
    \bar \phi &=&\frac{\phi_{09}+\phi_{14}}{2},\\
    A_{atm} &=& \frac{\phi_{09}-\phi_{14}}{2 \phi_{09}},\\
    \omega &=&\frac{2\pi}{11 \textrm{yr}}.
\end{eqnarray*}
The quantities $\phi_{09}$ and $\phi_{14}$ are the fluxes 
in the year of 2009 and 2014, corresponding to Solar min and Solar max, respectively. 

\subsection{Previous Experimental Methods and Results}
Over the past several decades, multiple experiments have searched for possible periodic signals in neutrino data. To this point, the eccentricity has been measured by Borexino, and the day-night modulation due to charged current interactions has been established by SK. An upper limit of the diurnal modulation has been set by Borexino. 

\par 
A typical data analysis procedure is to bin the observed data in time bins. Then, a Fourier transform is performed, a range of frequencies is scanned, and then peaks that correspond to possible periodic signals are found. The peaks are checked to determine if the power of the peak exceeds a certain threshold (false-alarm probability). The specific details of turning time-binned data into the periodogram, the error of the observed data, and the terminology describing the methods differ among the experimental analyses. 

\par In Table~\ref{tab:experiments}, we summarize the results from experiments at Homestake, Kamioka, Sudbury, Borexino, and IceCube. We refer to each paper for more details on their analysis methods. We characterize the experiments in terms of their run time $\T$, assumed time binning $\Delta t$, frequency range scanned, the periodic signal of interests, and the results. These summarized results provide a point of comparison for our future projections for dark matter detectors.

\section{Statistical Methods for Time-varying Analysis\label{sec:statistical}} This section establishes our statistical methods for detecting solar and atmospheric neutrinos, specifically describing our methods for constructing the likelihood function. In this case, the null hypothesis $H_0$ is defined as a signal with a constant rate in time, while the alternative, $H_1$, is defined as a signal whose rate of events varies in time with a period, $P$, and an amplitude, $A_d$.  

We  use the following models under $H_0$ and $H_1$ to formulate our likelihood ratio test. Let $n_i$ be the 
number of observed events in the time bin $(t_i, t_i+\Delta t)$. We  assume that 
\begin{eqnarray*}
    n_i\equiv 
    n(t_i)\sim \left\{
    \begin{array}{cr}
     {\rm Pois}(\lambda \Delta t) & \mbox{under } H_0,\\
      {\rm Pois}(\lambda_{i}) & \mbox{under } H_1,
    \end{array}\right.
    \label{eqn:lambda_model0}
\end{eqnarray*}
where the expected number of events in the  time bin $(t_i, t_i+\Delta t)$ under the assumption of $H_1$ is 
$$\lambda_{i} \equiv \lambda(t_{i}) =  \lambda \int^{t_i+\Delta t}_{t_i}\{1 + A \cos (\omega u) + B \sin (\omega u) \}du.$$
In the above formulation,  $\lambda$, $A$, and $B$ are unknown parameters, and $\omega$ is an unknown frequency.
In $\lambda_i$, 
the background signal is 
$\lambda \Delta t$ and the time-varying signal is $\lambda \int^{t_i+\Delta t}_{t_i}\{A \cos (\omega u) + B \sin (\omega u) \}du$. Note that $H_0$ is a special case of $H_1$ as we obtain the null model by setting $A=B=0$ in $\lambda_i$.

First, we assume that $\omega$ is known for a gentle exposition of the methodology. To test $H_0$ against $H_1$,  we use the generalized likelihood ratio (GLR) method. The term 
generalized is used when at least one of $H_0$ and $H_1$ is a composite hypothesis; in our case, both are composite hypotheses. 
\cite{2007:SNO} used the GLR method for identifying the periodicity of $^8$B solar
neutrino flux released by the SNO Collaboration. 
However, unlike \cite{2007:SNO}, we obtain the expected number of events within a time bin by integrating the time-varying instantaneous rate function, which is an appropriate method, especially when the bin width is not very small. Moreover, besides  applying to $^8$B solar neutrinos, we use this model for all components of the solar and atmospheric neutrino flux.  
The GLR statistic is 
\begin{equation}
    {\rm GLR} = \frac{{\rm max}_{ \lambda}\prod_{i=1}^{N_{t}} \frac{ \lambda^{n_{i}} e^{- \lambda}}{n_{i}!}}{ {\rm max}_{ \lambda, A, B}\prod_{i=1}^{N_{t}} \frac{\lambda_{i}^{n_{i}} e^{-\lambda_{i}}}{n_{i}!}}=\frac{{\rm max}_{ \lambda}\prod_{i=1}^{N_{t}}  
    \lambda^{n_{i}} \exp(- \lambda)}{ {\rm max}_{ \lambda, A, B}\prod_{i=1}^{N_{t}} \lambda_{i}^{n_{i}} \exp(-\lambda_{i})}. 
    \label{eqn: GLR}
\end{equation}
The numerator of (\ref{eqn: GLR}) is 
$$  
(\wh{\lambda}_0
\Delta t)^{\sum^{N_t}_{i=1}n_i}
\exp(-\wh{\lambda}_0\Delta tN_t),
$$
where $\wh{\lambda}_0=\sum^{N_t}_{i=1}n_i/N_t\Delta t$ denotes the maximum likelihood estimator of $\lambda$ under $H_0$.  
The denominator of (\ref{eqn: GLR})
 is 
 $$
\prod_{i=1}^{N_{t}} \wh\lambda_{i}^{n_{i}} \exp(-\wh\lambda_{i}), 
$$
where $\wh\lambda_i= \wh{\lambda}_1 \int^{t_i+\Delta t}_{t_i} \{1 + \wh{A} \cos (\omega u) + \wh{B} \sin (\omega u) \}du$ with $\wh{\lambda}_1$, $\wh A$ and $\wh B$ are the solution of the following gradient equations (obtained by differentiating the log-likelihood function under $H_1$)
\begin{eqnarray}
0&=&\sum_{i=1}^{N_{t}}\left[ \frac{n_i}{\lambda}- 
\int^{t_i+\Delta t}_{t_i}\{1+A\cos(\omega u)+ B\sin(\omega u)\}du
\right],\label{eq:g1}\\
    0 &= &\sum_{i=1}^{N_{t}} \left\{\frac{n_i \int^{t_i+\Delta t}_{t_i} \cos (\omega u)du}{
    \int^{t_i+\Delta t}_{t_i} \{    
    1 + A \cos (\omega t_{i}) + B \sin (\omega u)\}du}-  \lambda \int^{t_i+\Delta t}_{t_i}\cos (\omega u)du \right\},\label{eq:g2}\\
    0 &= &\sum_{i=1}^{N_{t}} \left\{\frac{n_i \int^{t_i+\Delta t}_{t_i} \sin (\omega u)du}{\int^{t_i+\Delta t}_{t_i} \{    
    1 + A \cos (\omega t_{i}) + B \sin (\omega u)\}du}-  \lambda \int^{t_i+\Delta t}_{t_i}\sin (\omega u)du \right\}. \label{eq:g3}
\end{eqnarray}
Equation (\ref{eq:g1}) yields 
$$\lambda= \frac{\sum^{N_t}_{j=1}n_j}{  
\sum^{N_t}_{j=1} \int^{t_j+\Delta t}_{t_j}\{ 1+ A\cos(\omega u)+ B\sin(\omega u)\}du},$$ and using this expression in 
(\ref{eq:g2}) and (\ref{eq:g3}) we obtain 

\begin{equation}
    \label{eqn:maxLH_AB}
    \begin{split}
    0 &= \sum_{i=1}^{N_{t}} 
    \left[\frac{n_i}{
    \int^{t_i+\Delta t}_{t_i}
    \{    
    1 + A \cos (\omega u) + B \sin (\omega u)\}du}- \frac{\sum^{N_t}_{j=1}n_j}{  
\sum^{N_t}_{j=1} \int^{t_j+\Delta t}_{t_j}\{ 1+ A\cos(\omega u)+ B\sin(\omega u)\}du}
\right] \int^{t_i+\Delta t}_{t_i}\cos (\omega v)dv  ,\\
    0 &= \sum_{i=1}^{N_{t}} 
   \left[\frac{n_i  }{
    \int^{t_i+\Delta t}_{t_i}
    \{
    1 + A \cos (\omega u) + B \sin (\omega u)\}du}- \frac{\sum^{N_t}_{j=1}n_j}{  
\sum^{N_t}_{j=1} \int^{t_j+\Delta t}_{t_j}\{ 1+ A\cos(\omega u)+ B\sin(\omega u)\}du} \right] \int^{t_i+\Delta t}_{t_i}
    \sin (\omega v) dv . 
    \end{split}
\end{equation}
To solve (\ref{eqn:maxLH_AB}),
we assume that data 
$n_i$'s and the frequency $\omega$ are known, and  replace
\mbox{$\int^{t_i+\Delta t}_{t_i}
    \sin (\omega v)dv$} by 
    \mbox{$[\cos(\omega t_i)-\cos\{\omega (t_i+\Delta t)\}]/\omega$}
and \mbox{$\int^{t_i+\Delta t}_{t_i}
    \cos (\omega v)dv$} by 
    \mbox{$[\sin\{\omega (t_i+\Delta t)\}-\sin(\omega t_i)]/\omega$}.
 The likelihood spectrum at $\omega$ is 
\begin{equation}
    \begin{split}
    S(\omega) = - \ln( {\rm GLR})
    =\left[\sum_{i=1}^{N_{t}} \left\{n_{i} \ln(\wh\lambda_{i}) - \wh\lambda_{i} \right\} - \ln(\wh \lambda_0\Delta t)\sum_{i=1}^{N_{t}} n_{i}  + N_t\wh  \lambda_0\Delta t \right], 
    \end{split}
    \label{eqn:S}
\end{equation}
where the expression of $\wh\lambda_i$ is given before equation (\ref{eq:g1}).
For a given value of $\omega$, this is the key statistic used for the GLR test. 

\par We can assume that the frequency $\omega$ is known or unknown when analyzing data. If $\omega$ is known and $N_t$ is large, then to test the hypotheses (for detecting the signal), one can reject $H_0$ when 
$2S(\omega)>\chi^2_{2, \alpha}$, where $\chi^2_{2, \alpha}$ denotes the upper $\alpha$ percentile of the Chi-square distribution with $2$ degrees of freedom. 
For an unknown $\omega$, we define the test statistic as 
\begin{eqnarray}
S_{\max}= \max_{\omega\in \Omega} S(\omega),\nonumber\label{eq:smax}
\end{eqnarray}
where $\Omega$ is set of set of $M$ distinct frequencies. 
In this case, we reject $H_0$ when 
$  S_{\max}>S_{\max, \alpha}$, where $S_{\max, \alpha}$ denotes the upper $\alpha$ percentile of the $S_{\max}$ distribution. 
Since the analytical form of the 
distribution is unknown, the percentile 
$S_{\max, \alpha}$
 is determined by a numerical procedure given in the next section. 

\par Alternative to the GLR method, we also consider the  Lomb-Scargle method \cite{1982Scargle}, hereafter referred to as LS. The  periodogram at a given frequency $\omega$ is 
\begin{equation*}
LS(\omega) = \frac{1}{2\sigma^2} \left( \frac{\left[\sum_{i=1}^{N_{t}} n_{i} \cos\{\omega (t_{i} -\tau)\}\right]^2 }{ \sum_{i=1}^{N_{t}} \cos^2\{\omega (t_{i} -\tau)\} } + \frac{\left[\sum_{i=1}^{N_{t}} n_{i} \sin\{\omega (t_{i} -\tau)\}\right]^2 }{ \sum_{i=1}^{N_{t}} \sin^2\{\omega (t_{i} -\tau)\} }\right)
\end{equation*}
with 
$\tau = (1/2 \omega) \tan^{-1} \left\{ \sum_{i=1}^{N_{t}} \sin (2 \omega t_{i})/\sum_{i=1}^{N_{t}} \cos (2\omega t_{i})\right\}
$
and 
$\sigma^2 = \sum_{i=1}^{N_{t}} n_{i}^2 /(N_{t} - 1)$.
Like the GLR case, for known $\omega$, we reject $H_0$ if $2L(\omega)>\chi^2_{2, \alpha}$ and 
for the unknown $\omega$ we reject $H_0$ if $L_{\max} =\max_{\omega\in \Omega}L(\omega)>L_{\max, \alpha}$. 
This $L_{\max, \alpha}$ is the upper $\alpha$ percentile point of $L_{\max}$. 
This percentile (or the critical value of this test) can be calculated analytically or numerically. Ref.~\cite{2007:SNO} approximated the distribution of $L_{\max}$
by the following probability density function
\begin{equation}
    f_{X}(x|\nu) = \nu (1-e^{-x})^{\nu-1} e^{-x},  x>0,
       \label{eqn:pdf_M}
\end{equation}
where $\nu$ is a function of the number of scanned frequencies in $\Omega$. Following \citep{2003PhRvD..68i2002Y}, we take $\nu=\M_{scan}$, and the details on how to calculate $\M_{scan}$ is given in the next section. To be specific,  Equation (\ref{eqn:pdf_M}) is the density of the maximum of $\nu$ independent standard exponential 
($\chi^2_2$ /2) random variables. Therefore, we can determine the upper $\alpha$ percentile of $LS_{\max}$ from 
the density given in Equation (\ref{eqn:pdf_M}). 
The upper $\alpha$ percentile, denoted by $LS_{\max, \alpha, 1}$ is $- \log \{ 1-(1-\alpha)^{1/\nu}\}$. We 
use $LS_{\max, \alpha, 2}$ to denote the  percentile determined by a simulation technique.

 \section{Synthetic data simulation strategy}\label{sec:data_analysis}
 \subsection{General simulation strategy}
 We generated data under $H_0$ and $H_1$. 
The counts $n_i$ over the time bin $[t_i, t_i+\Delta t)$
follow
\begin{equation}\nonumber
    \begin{array}{lr}
     {\rm Pois}\{({\cal R}_s + {\cal R}_b){\cal D} \Delta t \} &\mbox{under }H_0,\\
    {\rm Pois}\{{\cal R}_{b}{\cal D} \Delta t+ 
 {\cal R}_{s}{\cal D}(\Delta t + A_d \int^{t_i+\Delta t}_{t_i} 
  \sin\left(\frac{2\pi}{P} u - \Psi \right)  du )\} &\mbox{under }H_1.
    \end{array}
\end{equation}

 Here ${\cal R}_{s}$ is a constant signal event rate and ${\cal R}_{b}$ is a constant rate for the background noise. Here, we have chosen not to bin in recoil energy, so ${\cal R} = {\cal R}_{s}+{\cal R}_{b}$ denotes the event rate at any given time integrated over the entire recoil energy range and summing over all flux components. For analyzing (testing of hypotheses) simulated data, we employed the GLR and LS methods.

Note that we use a value of $\omega$ for data simulation. But at the analyses stage, $\omega$ is assumed to be once known and then an unknown parameter.  
For unknown $\omega$, the statistic $S_{\max}$ and the critical value of $S_{\max}$ are determined as follows. 
We take $\Omega$ as a set of $N_{\rm entire} = n_{o}{\cal T}/\Delta t$ 
evenly-spaced frequencies on the entire range for the scanned frequencies $[f_{min}, 1/\Delta t]$  for a given run time ${\cal T}$, width of the time interval $\Delta t$, and over-sample factor $n_o$~\citep{2018VanderPlas}, so 
$\Omega=\{\omega_1, \dots, \omega_{N_{entire}}\}$. We set $n_o=10$ for the Monte Carlo simulation. We use the same $\Omega$ for both  $S_{\max}$ and $L_{\max}$. 
Another concept needed for deriving the null distribution of $L_{\max}$ is the effective number of scanned frequencies over $[f_{min}, 1/\Delta t]$. It is ${\cal M}_{entire}= N_{entire}/f_{adj}$, where  $f_{adj}$ is an empirical result for the scanning density. 
Note that we are interested in  the frequency range $[f_{min}, f_{max}]$ with $f_{max} = {\rm min}(1/\Delta t, 1 \, \textrm{day}^{-1})$, and $f_{min} = 1/365.25 \, {\rm day}^{-1}$ for $P$ = 1 yr or 1 day, $f_{min} = 1/11 \, {\rm yrs}^{-1}$ for $P$ = 11 yr. Then we over scan $N_{entire}\times f_{scan}/f_{entire}$ on $[f_{min}, f_{max}]$ where $f_{entire} =1/\Delta t - f_{min}$ and $f_{scan} =f_{max} - f_{min}$. Specifically,  to be consistent with the frequency range of interest, $f_{adj}$ for different $\Delta t$ and $\T$ are showed in Appendix~\ref{sec:appendix}. The number of independent frequencies scanned in a smaller frequency range \citep{2003PhRvD..68i2002Y} is 
${\cal M}_{scan} = {\cal M}_{entire} \times f_{scan}/f_{entire}$, and this $\M_{scan}$ is used to approximate the distribution of $L_{\max}$ given in (\ref{eqn:pdf_M}). 

\subsubsection{ Critical value $S_{\rm max, \alpha}$ from the distribution of $S_{\max}$}
We found that the empirical density of $S_{\max}$ is somewhat close to (\ref{eqn:pdf_M}) for some choice of $\nu$, which need not be the same as the number of frequencies of $\Omega$. 
However, it isn't easy to prove that the density of $S_{\max}$ has the form (\ref{eqn:pdf_M}) because
1) for any two arbitrary frequencies  $\omega$ and $\omega^{'}$, both $S(\omega)$ and $S(\omega^{'})$ are calculated on the same observed data, just at two different frequencies, so they need not be independent, and 2) finding the distribution of the maximum of dependent $\chi^2$ random variables is not an easy task. Therefore, the threshold $S_{\max, \alpha}$ estimated after fitting 
(\ref{eqn:pdf_M}) to the $S_{\max}$ data, will be an approximation of the upper $\alpha$ percentile of $S_{\max}$'s distribution. 
For each $5000$ simulated dataset under $H_0$, we computed $S_{\max}$, which helps to obtain the empirical distribution of $S_{\rm max}$. 
We then fit model (\ref{eqn:pdf_M}) to the  $S_{\rm max}$ values denoted by $\{ S_{{\rm max}, r}, r=1, \dots, 5000\}$ and obtain the maximum likelihood estimates (MLE) of $\nu$ by maximizing the following log-likelihood function: 
\begin{eqnarray*}
\sum^{5000}_{r=1}[\log(\nu)+(\nu-1)\log\{1- \exp(-S_{{\rm max}, r}) \}- S_{{\rm max}, r}].  
\end{eqnarray*}
Let us denote the MLE of $\nu$ by $\wh\nu$. Since the cumulative distribution function of the probability density  (\ref{eqn:pdf_M}) is 
\mbox{$F(h)= \{1-\exp(-h)\}^{\nu}$}, we set \mbox{$\{1-\exp(- S_{\rm max, \alpha})\}^{\wh\nu}=1-\alpha$}, and solving this equation we obtain the  estimate $S_{\max, \alpha, 1}=- \log \{ 1-(1-\alpha)^{1/\wh\nu}\}$. 
Next, we use the upper 
$\alpha$th percentile of the 5000 $S_{\rm max}$ values as the second  estimate of $S_{\max, \alpha}$, call it $S_{\max, \alpha, 2}$. 

\subsubsection{ Results between GLR and LS methods under $H_0$ }
In our simulation, we set $\alpha = 0.1$. The scanned frequencies were taken between [$f_{min}$, $f_{max}$] with a bin-width of  $\Delta t$, run-time 
${\cal T}$, and the detector size ${\cal D} = 100$ ton. 
Computation of the threshold under the Lomb-Scargle method was much faster as it did not involve any equation solving, even when the number of frequencies was in the order of $10^5$. In contrast, the GLR-based approach was time-consuming. Specifically, the computing time of GLR was proportional to the number of scanned frequencies and the length of $n_i$. We used scipy.root~\cite{2020SciPy-NMeth} 
and astropy.timeseries.LombScargle~\cite{astropy:2013}
of Python for GLR and LS calculations, respectively. 
Table~\ref{tab: M} contains the critical values for the GLR and LS methods when ${\cal T}$ = 10 yrs. 

\begin{table}
\caption{
This table shows the threshold values of the test statistic derived from 1) the Lomb-Scargle periodogram and 2) the GLR, for different 
$\Delta t$, $\mathcal{T}$, ${\mathcal M}_{scan}$ and when the detector's size is assumed to be $\mathcal{D}=100$ tons, \mbox{${\cal R}_{s} = 100$ ton$^{-1}$ yr$^{-1}$} and \mbox{${\cal R}_{b} = 0$ ton$^{-1}$ yr$^{-1}$}, and  $\alpha=10\%$.  
The data are generated under $H_0$. Under each method two estimates of threshold are presented, a) model based corresponds to the one with suffix 1 and b) empirical distribution based, corresponds to the one with suffix 2.   All results are based on simulation with 5000 replications.
}
\begin{tabular}
{ r r r r |rr r |r r r}
\hline
${\cal T}$ & $\Delta t$ &  $\D$ & average number & \multicolumn{3}{c|}{LS} &\multicolumn{3}{c}{GLR}\\

years& days & tons &of events per bin  &${\mathcal M}_{scan}$ & 
$LS_{\max, \alpha, 1}$& $LS_{\max, \alpha, 2}$ & $\wh\nu$  & $S_{\max, \alpha, 1}$& $S_{\max, \alpha, 2}$ 
 \\
\hline
10 & 30 & 100 & 821 & 202   & 7.56	&  7.24  & 133  & 7.14 & 7.39
\\
10& 10 & 100 & 273 & 608 & 8.66 &  8.4 &  463  & 8.39 & 8.49
\\
10& 5 &100 & 136 & 1217  & 9.35 & 9.23  & 946  & 9.1 & 9.26
\\
\hline
$1/6$ & 0.2 & 100  & 5  &  404   &  8.25  & 8.05  & 329  & 8.05  & 8.14\\
$1/6$& 0.2 & 500  & 27 & 404 & 8.25 & 8.03 &  328 & 8.04 & 8.16\\
$1/6$& 0.2 & 2000 & 109 & 404 & 8.25 & 8.12 & 325 &8.03 & 8.22\\
\hline
\end{tabular}
\label{tab: M} 
\vskip 2mm 
\end{table}

\subsubsection{Simulation procedure under $H_1$}

The power of the test is the probability of rejecting $H_0$ when data are generated under the alternative hypothesis $H_1$. The simulated data under $H_1$ is a Poisson distribution with a time-varying rate determined by a given amplitude $A_d$ and period $P$. We choose $\Psi = P/4$ such that the simulated time variation starts at the extreme value.

\par We are mainly interested in the periodic signals in the neutrino flux from the Sun with a period of \mbox{$P=1$ day} (due to Earth rotating itself), \mbox{$P=365.25$ days} (due to Earth orbiting around the Sun), and atmosphere \mbox{$P=11$ yrs} (due to the 11 yr solar cycle, the changing strength of the solar wind). All three periods should in principle be observable. In the data generation process, we consider $P=1$ day and 365.25 days with \mbox{$\Delta t <  1$ day}; \mbox{$P= 365.25$ days} with \mbox{$1$ day $\leq \Delta t< 365.25$ days}; \mbox{$P= 11$ yrs} when \mbox{$\Delta t <  1$ yr} and \mbox{$\T \geq 11$ yrs}. 

When $n_i\stackrel{ind}{\sim}{\rm Poisson}(\lambda_i), i=1, \dots, N_t$ , the  power of the tests is calculated under two scenarios

\begin{enumerate}

\item 
\underline{Known period P}: Under such cases when the periods are known, instead of searching for all possible physical and theoretical peaks by finding a peak as is being done in actual experiments, we calculate the statistics $S(\omega_{known})$ and $LS(\omega_{known})$ at that known $\omega_{known}=2\pi/P$, instead of estimating $\omega$. 
The powers of the tests are 
\begin{eqnarray*}
        {\rm Power}= \left\{ \begin{array}{cr}
        \pr\{ 2S(\omega_{known})> \chi^2_{2, \alpha}\}
        , & \mbox{for GLR, } \\
        \pr\{ 2LS(\omega_{known})> \chi^2_{2, \alpha}\}
        ,& \mbox{for LS.} 
        \end{array}\right.
        \label{eqn:power_knownP}
        \end{eqnarray*}

\item 
\underline{Unknown period P}: Experimentally, possible periodic signals are searched by finding peaks over a wide frequency range. The scanned frequency range and frequency number are mentioned in the third paragraph of this section. In this case, $S_{max}$ may be located at the alias of the input frequency instead of at the input frequency.
    The powers of the tests are
        \begin{eqnarray*}
        {\rm Power}= \left\{ \begin{array}{cr}
        \pr( S_{\max}> S_{\max, \alpha}), & \mbox{for GLR}, \\
        \pr( LS_{\max}> LS_{\max, \alpha}), & \mbox{for LS}. 
        \end{array}\right.
        \label{eqn:power_unknownP}
        \end{eqnarray*}

\end{enumerate}

We define $\beta = 1 - {\rm Power}$ as the probability of failing to reject the null hypothesis when the alternative is true (i.e.,  the type II error). For the empirical power (power obtained via simulation), we first fix  $(\R_b, \R_s, \D, \T, P, \Psi, \Delta t)$. Then change $A_d$
 over a grid of values. For every choice of 
 $(\R_b, \R_s, \D, \T, P, \Psi, \Delta t, A_d)$, we generate Poisson data $\{n_1, \dots, n_{N_t}\}$ under $H_1$.
We calculate $S(\omega_{known})$ or $LS(\omega_{known})$ or $S_{max}$ or $LS_{max}$ for this simulated data. We repeat this procedure $5000$ times, then calculate the proportion of times these quantities is larger than $\chi^2_{2, \alpha}$ or $S_{\max, \alpha}$ or $LS_{\max, \alpha}$ accordingly. This proportion is an estimate of the power (Eqn~\ref{eqn:power_knownP}). This procedure is repeated for every combination of the parameters. Next we draw the power curves against $A_d$ for a fixed value of $(\R_b, \R_s, \D, \T, P, \Psi, \Delta t)$ and we denote the amplitude $A_d$  as $A_{d, \alpha, \beta}$ when the probability of type-I error rate of a hypothesis test being $\alpha$ and type-II error being $\beta$.

\begin{figure*}[!htbp]
    \includegraphics[width = 0.9\textwidth ]{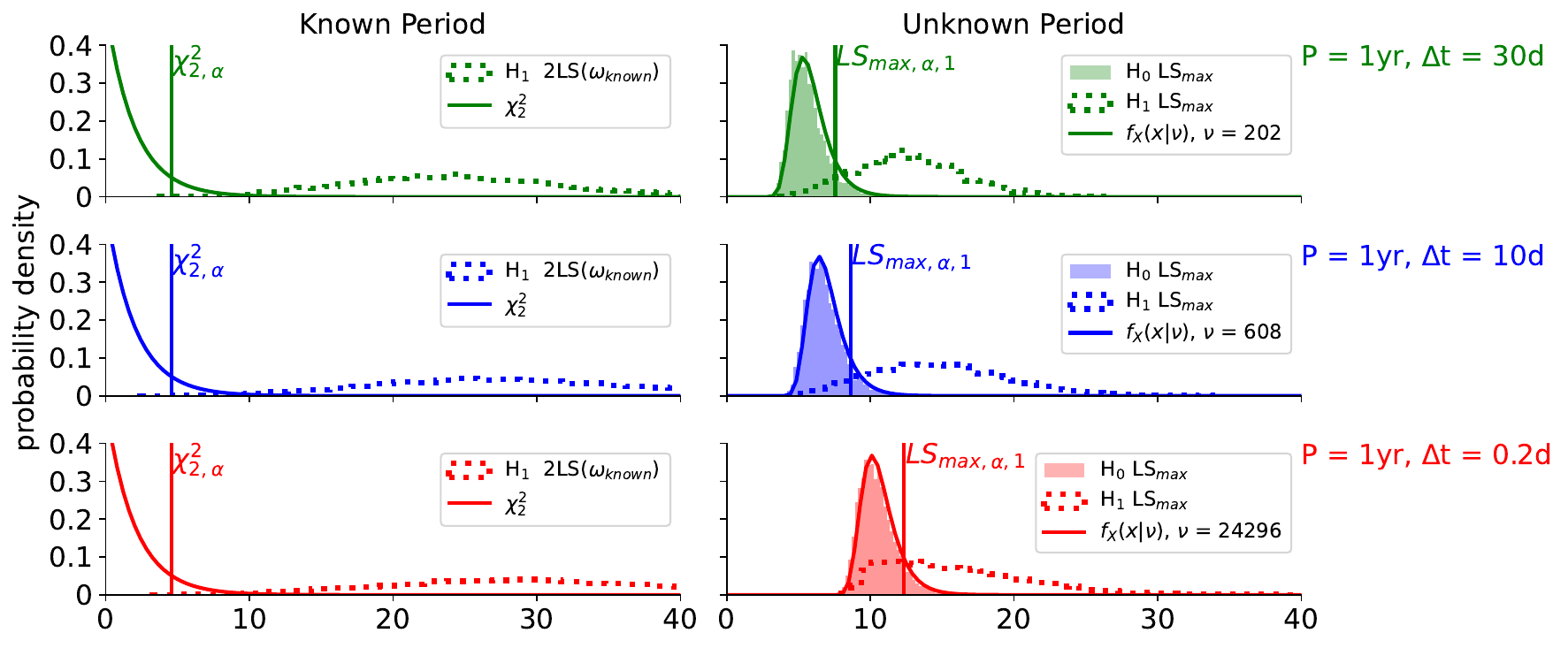}
    \caption{Example of histograms for the assumptions of Known Periods (left)  and Unknown Periods (right) under the null hypothesis $H_0$ and the alternative hypothesis $H_1$. Assumed parameters are runtime ${\cal T}$ = 10 yrs, time binnings $\Delta t= 0.2, 10, 30$ days, period of $P = 365.25$ days, amplitude $A_{d} = 0.024$, \mbox{${\cal R}_{s} = 100$ ton$^{-1}$ yr$^{-1}$}, and no experimental backgrounds \mbox{${\cal R}_{b} = 0 $ton$^{-1}$ yr$^{-1}$}. {\it Left column}: The solid curve is the $\chi^2_2$ distribution, the vertical line is the critical value $\chi^2_{2, \alpha}$ with $\alpha$ = 0.1. The dotted histogram in each panel is the distribution of $5000$ simulated values of $2LS_{max}$ under the alternative hypothesis. {\it Right column}: The filled histogram in each panel is the empirical distribution of $5000$ simulated values of $LS_{\rm max}$ under the null hypothesis $H_0$ using the LS method as described in the text. The solid curve is $f_X(x|\nu)$ with $\nu = {\cal M}_{scan}$, the vertical line is the critical value $LS_{\max, \alpha, 1}$ with $\alpha$ = 0.1. The dotted histogram in each panel is the distribution of $5000$ simulated values of $LS(\omega_{known})$ under the alternative hypothesis with $\omega_{known} = 2\pi/P$ at the detection period $P= 365.25$ days.
    }
    \label{fig:LS_eg}
\end{figure*}

\par~Figure~\ref{fig:LS_eg} shows an example when the detecting period $P = 365.25$ days using the run time ${\cal T} = 10$ yrs, and binned in \mbox{$\Delta t = 0.2, 10, 30$ days}. The left histograms are under the ``Known Period" condition, where $H_0$ is the $\chi^2_2$ distribution and $H_1$ is $2LS(\omega_{known})$. The right histograms are under the ``Unknown Period" condition , where $H_0$ and $H_1$ are $LS_{max}$. The amplitude for $H_1$ is $A_d = 0.024$. 
The vertical line in each panel is the critical value $\chi^2_{2, \alpha}$ (left column) and \mbox{$LS_{\max, \alpha, 1}$ } (right column) when \mbox{$\alpha = 0.1$}. The power $(1-\beta)$ is the proportion of $2LS(\omega_{known})$ (or $LS_{\max}$) greater than the threshold $\chi^2_{2, \alpha}$ (or $LS_{\max, \alpha, 1}$) out of the 5000 realizations under $H_1$, where $\omega_{known} = 2\pi/P$.

\subsubsection{Comparing $A_{d, \alpha, \beta}$ between GLR and LS methods}

\begin{figure}[!htbp]
    \includegraphics[width = 0.55\textwidth ]{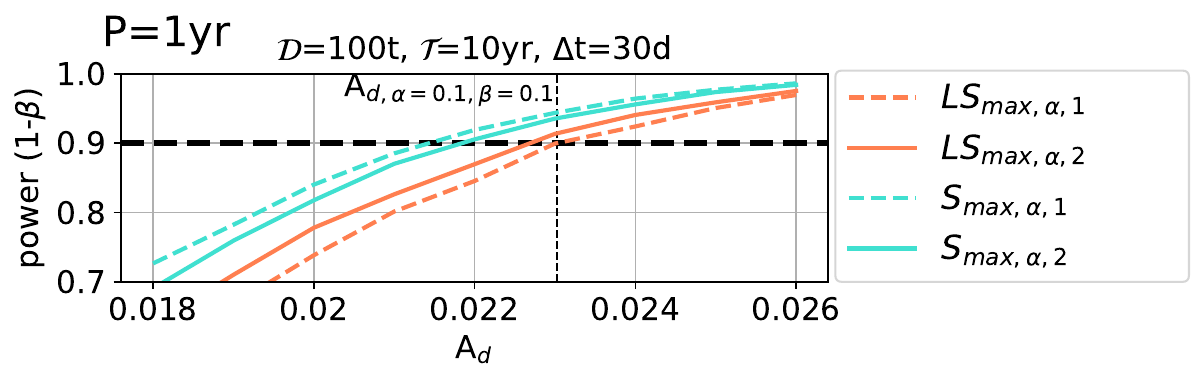}
    \caption{Power curves (the probability of rejecting $H_0$) for the GLR and Lomb-Scargle methods when $\alpha = 0.1$ and $P = 1$ yr and under the ``Unknown Period" scenario. The critical values are $LS_{\max, \alpha, 1}$, $LS_{\max, \alpha, 2}$  for LS and $S_{\max, \alpha, 1}$ and $S_{\max, \alpha, 2}$  for GLR. The vertical line is $A_{d, \alpha = 0.1, \beta = 0.1}$ under the LS method and when the critical value is  $LS_{\max, \alpha, 1}$.
    }
    \label{fig:cfLSGLR_1yr1d}
\end{figure}

For a fixed value of $(\R_b, \R_s, \D, \T, P, \Psi, \Delta t)$, we want to compare $A_{d, \alpha, \beta}$ between LS and GLR methods, under ``Known Period" and ``Unknown Period" scenarios. We also test the performance for a small time bin and a small number of events~\cite{Borexino:2022khe}. 
For these purposes, we set \mbox{${\cal R}_{s} = 100$ ton$^{-1}$ yr$^{-1}$} and \mbox{$\R_b = 0$ ton$^{-1}$ yr$^{-1}$}. We consider  \mbox{$\T = 1/6$ yrs} with \mbox{$\Delta t = 0.2$ days}, and \mbox{$\T = 10$ yrs} with \mbox{$\Delta t = 10, 30$ days}, and control the number of events in each time bin by changing $\D$. One example is shown in Figure~\ref{fig:cfLSGLR_1yr1d}, and results are shown in Table~\ref{tab: cf_LSGLR}. 
Under the ``Known Period" scenarios, the results from GLR and LS are similar. 
However, under the unknown period scenario, the power of GLR is generally better than the LS method, especially when the number of events in time bins is small. Both methods become equivalent in power as the number of events in each bin increases.
We adopt LS in the rest of the analyses as it is computationally more efficient than GLR, especially for small time bin $\Delta t$ and long run time $\T$.

\begin{table}[!htbp]
\caption{This table shows the amplitude $A_{d, \alpha, \beta}$ when $\alpha$ = 0.1, $\beta$ = 0.1 for scenarios 1) Known Period and 2) Unknown Period, and methods 1) the Lomb-Scargle periodogram and 2) the GLR, for different $\D$, $\T$, $\Delta t$ and when \mbox{$\R_s = 100$ ton$^{-1}$ yr$^{-1}$}, \mbox{$\R_b = 0$ ton$^{-1}$ yr$^{-1}$}. Under the Known Period scenario, the critical value is $\chi^2_{2,\alpha}$ for LS and GLR. Under the Unknown Period scenario, the critical values are $LS_{\max, \alpha, 1}$, $LS_{\max, \alpha, 2}$  for LS and $S_{\max, \alpha, 1}$ and $S_{\max, \alpha, 2}$  for GLR.
}
\begin{tabular}
{ r r r r r | r r | r r r r }
\hline
\multirow{2}{*}{${\cal T}$} & \multirow{2}{*}{$\Delta t$}&  \multirow{2}{*}{Period} & \multirow{2}{*}{$\D$} & average  &\multicolumn{6}{c}{$A_{d, \alpha, \beta}$ }\\
\cline{6-11}
&&  &   &  number of & \multicolumn{2}{c|}{``Known Period " scenario} &\multicolumn{4}{c}{``Unknown Period" scenario} \\
\cline{6-11}
years & days & days & tons &     events per bin &LS & GLR &$LS_{\max, \alpha, 1}$& $LS_{\max, \alpha, 2}$ & $S_{\max, \alpha, 1}$& $S_{\max, \alpha, 2}$\\
\hline
10 & 30 & 365.25 & 100  & 821 & 0.015  & 0.015 & 0.023 & 0.023 & 0.021& 0.022\\
10 & 10 & 365.25 & 100 & 273 & 0.015  & 0.015 & 0.023 & 0.023 & 0.023 & 0.023\\
\hline
$1/6$ & 0.2 & 1 & 100  & 5 & 0.12 & 0.12 & 0.195 & 0.193 & 0.188 & 0.189  \\
$1/6$ & 0.2 & 1 &  500  & 27  &0.055 & 0.054 & 0.087 & 0.086 & 0.084 & 0.084\\
$1/6$ & 0.2 & 1 & 2000 & 109 & 0.027 & 0.027 & 0.043 & 0.043& 0.042 & 0.042\\    
\hline
\end{tabular}
\label{tab: cf_LSGLR} 
\vskip 2mm 
\end{table}

\subsection{Estimating $A_{d,\alpha,\beta}$ for arbitrary ($\R_s$, $\R_b$, $\D$) and fixed ($\T$,$\Delta t$, $P$) via the signal-to-noise ratio}
In the previous section, $A_{d, \alpha, \beta}$ is obtained via a simulation process for a given $(\R_b, \R_s, \D, \T, P, \Psi, \Delta t)$. This process is time intensive. It is desirable to obtain an analytical expression for the variation of $A_{d,\alpha,\beta}$ when ($\R_{s}$, $\R_{b}$, $\D$) take on different values. The signal rate $\R_{s}$ depends on flux uncertainty, $\sin\theta_{w}^2$, nuclei, physical process, energy threshold $E_{thrd}$, detector efficiency $\epsilon(E_r)$ and resolution $\sigma(E_r)$. The background rate $\R_{b}$ depends on the background depletion and physical process to generate the background. The detector size $\D$ depends on the experiment. In this section, after obtaining $A_{d, \alpha, \beta}$ for $(\R_b, \R_s, \D, \T, P, \Psi, \Delta t)$, we find an efficient way to obtain $A^{'}_{d, \alpha, \beta}$ for $(\R^{'}_b, \R^{'}_s, \D^{'}, \T, P, \Psi, \Delta t)$  so that the test maintains the same power.

 We set $\alpha = 0.1$ and $(1-\beta) = 0.9$ for the ``90\% detection'', then the corresponding detectable amplitude is denoted by $A_{d, \alpha = 0.1, \beta = 0.1}$. 
The signal for the detectable amplitude $A_{d,\alpha,\beta}$ in the time bin $[t_i, t_{i}+\Delta t]$ under $H_1$ is
\mbox{$\R_s \D A_{d,\alpha,\beta} \int^{t_i+\Delta t}_{t_i} \sin$ $\left(2\pi u/P\right)du$}.
Here $\alpha$, $\beta$, $P$, $\Delta t$, and $\cal T$ are assumed to be fixed (not changing). 
Also, the background is $\sqrt{(\R_s + \R_b)\D \Delta t}$, representing the Poisson standard deviation under the null model for the number of occurrences in one interval. Since the time-dependent parameters ($\T$, $\Delta t$, $P$) are fixed, the critical value stays the same during the scaling. In addition, the signal function deviated from the background on interval $[t_i, t_i+ \Delta t]$ is \mbox{$A_{d,\alpha,\beta} \R_s  \D   \int^{t_i+\Delta t}_{t_i} \sin\left(2\pi u/P\right) du$}. Component \mbox{$\int^{t_i+\Delta t}_{t_i} \sin\left(2\pi u/P\right) du$} determines the shape of the signal on $[0, T]$. When time-dependent parameters ($\T$, $\Delta t$, $P$) are fixed, the shape of this signal function over $[0, \T]$ may remain the same, but its magnitude may be different, depending on the parameters ($A_{d,\alpha,\beta}$, $\R_s$, $\D$). In this case, the power of the tests only depends on the signal's magnitude, because the shape's effect is the same. 
The squared signal-to-noise ratio ($SNR^2$) is 
\begin{equation}\nonumber
      \left.SNR^2 \right\vert_{\R_s, \R_b, \D} = A_{d,\alpha,\beta}^2 \times \R_s \times \D \times \frac{1}{(1+ \R_b/\R_s)} \times F(P, \T, \Delta t),
\end{equation}
where $F(P, \T, \Delta t)$ is some suitable effect due to the shape on the squared signal-to-noise ratio. 
Note that the 
signal-to-noise ratio for amplitude $A_{d,\alpha,\beta}$ when
there is no constant background  ($\R_s, \R_b=0, \D$), 
 is 
$$\left.SNR^2 \right\vert_{\R_s, \R_b=0, \D}  =A_{d,\alpha,\beta}^2 \times \R_s \times \D \times  
F(P, \T, \Delta t).$$ 
 Therefore, for any $(\R_s, \R_b, \D)$, 
\begin{eqnarray}\label{eq:1001}
SNR^2 \vert_{\R_s, \R_b =0, \D} = SNR^2\vert_{\R_s, \R_b >0, \D}  \times \left(1 + \frac{\R_b}{\R_s}\right).
\end{eqnarray}
 Next, 
\begin{equation}
\begin{split}
    \left.SNR^2 \right\vert_{\R_s, \R_b=0, \D} &=A_{d,\alpha,\beta}^2 \times \R_s \times \D \times  F(P, \T, \Delta t)\\
    &=A_{d,\alpha,\beta}^2 \times \R_s \left(\frac{\R_s^{'}}{\R_s^{'}}\right) \times \D \left(\frac{\D^{'}}{\D^{'}}\right) \times F(P, \T, \Delta t) \\
    &= \left(A_{d,\alpha,\beta} \sqrt{\frac{\R_s}{\R_s^{'}}} \sqrt{\frac{\D}{\D^{'}}}  \right)^{2} \times \R_s^{'} \times \D^{'} \times F(P, \T, \Delta t). 
    \end{split}\label{eq:1003}
\end{equation}
Now, suppose that 
for another set of values of $(\R^{'}_s, \R^{'}_b=0, \D^{'})$ the signal-to-noise ratio is the same as of $(\R_s, \R_b=0, \D)$ while all other parameters except the amplitude are fixed for both cases, i.e., 
\begin{equation}\label{eq:1002}
\left.SNR^2 \right\vert_{\R^{'}_s, \R^{'}_b=0, \D^{'}}=A_{d,\alpha,\beta}^{'2} \times \R^{'}_s \times \D^{'} \times F(P, \T, \Delta t)= 
\left.SNR^2\right\vert_{\R_s, \R_b=0, \D}.
\end{equation}

Now, using (\ref{eq:1001}) and (\ref{eq:1002}) we obtain 
\begin{eqnarray}
\left. SNR^2 \right\vert_{\R^{'}_s, \R^{'}_b >0, \D^{'}}  \times \left(1 + \frac{\R^{'}_b}{\R^{'}_s}\right)
&=&
\left.SNR^2\right\vert_{\R^{'}_s, \R^{'}_b =0, \D^{'}} =
\left.SNR^2 \right\vert_{\R_s, \R_b =0, \D},\label{eq:1051}\end{eqnarray}
and using (\ref{eq:1003}) 
in (\ref{eq:1051}) we have 
\begin{eqnarray*}
(A^{'}_{d,\alpha,\beta})^2 \times \R^{'}_s \times \D^{'} \times \frac{1}{(1+ \R^{'}_b/\R^{'}_s)} \times F(P, \T, \Delta t)&=& 
\left(A_{d,\alpha,\beta} \sqrt{\frac{\R_s}{\R_s^{'}}} \sqrt{\frac{\D}{\D^{'}}}  \right)^{2} \times \R_s^{'} \times \D^{'} \times F(P, \T, \Delta t),
\end{eqnarray*}
which yields the amplitude 
for other setting $(\R^{'}_s, \R^{'}_b=0,
\D^{'})$
\begin{eqnarray}
A_{d,\alpha,\beta}^{'} =A_{d,\alpha,\beta} \times \sqrt{\frac{{\cal R}_s}{{\cal R}_{s}^{'}}} \sqrt{\frac{{\cal D}}{{\cal D}^{'}}} \times  \sqrt{1+\frac{{\cal R}_{b}^{'}}{{\cal R}_{s}^{'}}}. 
    \label{eqn:1004}
\end{eqnarray}
As an example, 
for $\T = 10$ yrs, $\Delta t = 10$ days, $P$ = 365.25 days, \mbox{$\R_s = 100$ ton$^{-1}$ yr$^{-1}$}, \mbox{$\R_b = 0$ ton$^{-1}$ yr$^{-1}$}, \mbox{$\D = 100$ tons}, the detectable amplitude with a power of $90\%$ and a level of significance $10\%$ is \mbox{$A_{d, \alpha = 0.1, \beta = 0.1} = 
0.02357377$} (results in Figure~\ref{fig:cfLSGLR_1yr1d}).
Using Equation (\ref{eqn:1004}), we can now obtain the detectable amplitude 
$A_{d, \alpha = 0.1, \beta = 0.1}^{'} = 0.1292$ for 
\mbox{$\D^{'} = 100$ tons}, 
\mbox{$\R_s^{'} = 474.367$ ton$^{-1}$ yr$^{-1}$}
(ES Xenon resolution +efficiency) and \mbox{$\R_b^{'} = 50060.5$ ton$^{-1}$ yr$^{-1}$} (all backgrounds). 
This calculation is done without running a time-consuming simulation.
The detectable amplitudes for other possible values of $\R_s^{'}$ and 
$\R_b^{'}$ are given in the ``signal components" and ``background components" column in Table~\ref{tab:component_ES}~\ref{tab:component_CEVENS}.
Moreover, when the eccentricity $A^{'}_{d,\alpha = 0.1,\beta = 0.1}= A_{ecc} =0.03342$, using Equation (\ref{eqn:1004})
we obtain the detector's size $\D^{'}$ 
\begin{eqnarray}
     \D^{'} = \D 
     \left( A_{d,\alpha = 0.1,\beta = 0.1}
     \times \frac{\sqrt{1+\R_b^{'}/\R_s^{'}}}{A_{ecc}} 
     \right)^2  
     \left (\frac{\R_s}{\R_s^{'}}  \right)
     \nonumber
\end{eqnarray}
interpreted as the detector size required for the eccentricity to be $90\%$ detected.

Figures \ref{fig:ERamp_nubb} and \ref{fig:NRamp_thrdeff} show 
the amplitudes for different combinations of 
 $(\T, \Delta t, P, \alpha, \beta)$. 
 For example, running simulations we  obtain $A_{d,\alpha=0.1,\beta=0.1} = 0.0291161$, $A_{d,\alpha=0.05,\beta=0.1} = 0.03$ for $\T = 10$ yrs, $\Delta t = 0.2$ days, $P = 1$ day, \mbox{$\R_s = 100$ ton$^{-1}$ yr$^{-1}$}, \mbox{$\R_b = 0$ ton$^{-1}$ yr$^{-1}$}, \mbox{$\D = 100$ tons}. Then using Equation (\ref{eqn:1004}) we obtain $A_{d,\alpha=0.1,\beta=0.1}^{'}$, $A_{d,\alpha=0.05,\beta=0.1}^{'}$ for the same $\T$, $\Delta t$, $P$, and different choices of $(\R_s^{'}, \R_b^{'}, \D^{'})$. We also consider the cases, ($\T = 10$ yrs, $\Delta t = 0.2$ days, $P = 1$ day), 
 ($\T = 10$ yrs, $\Delta t = 10$ days, $P = 1$ yr) and ($\T = 10$ yrs, $\Delta t = 30$ days, $P = 1$ yr).

\subsection{Estimating $A_{d,\alpha,\beta}$ to have the required power with different $\Delta t$ and ${\cal T}$}
We now fix \mbox{$\R_s= 100$ ton$^{-1}$ yr$^{-1}$}, \mbox{$\R_b = 0$ ton$^{-1}$ yr$^{-1}$}, \mbox{$\D =  100$ tons} and explore the relation between the amplitude $A_{d, \alpha, \beta}$ and the time-dependent parameters ($\T, \Delta t, P$). For a given $(\T, \Delta t)$, we scan a series of frequencies ($P$) to obtain $A_{d,\alpha,\beta, P}$ for each $P$. Then we find the $P$ range where $A_{d, \alpha,\beta}$ is roughly constant $A_{d, \alpha, \beta, P} \approx \overline{A}_{d, \alpha,\beta}$ where we use $\overline{A}_{d,\alpha,\beta}$ to denote the average of the amplitudes over different $P$s.

The simulations show how different $T$ and $\Delta t$ affect the strength of signal ${A}_{d, \alpha,\beta}$ to maintain $90\%$ power of the test. 
With the requirement that  $P\gg \Delta t$, 
$\int^{t_i+\Delta t}_{t_i}
    \sin (\omega v)dv \approx \Delta t \sin(\omega t_i)$.
 Therefore, when $\R_b = 0$, the signal noise ratio 
$SNR^2_{\R_s, \R_b=0, \D}$  is approximated by ${A}_{d,\alpha,\beta}^2 \times \R_s \times \D \times \Delta t \times F_{sin}(\T, P)$, where $F_{sin}(\T, P)$ represents the effect on the signal-to-noise ratio due to different parameters $T$, and $P$, considering a signal with the shape of $sin(\omega t)$. The sample size of observations is $N_t = T/\Delta t$. The sample can differ with $\T$ and $\Delta t$. When all other parameters are fixed, but $\Delta t$ decreases, the signal-to-noise ratio decreases, while the sample size $N_t$ increases. Consequently, the expected power shall remain the same.

\begin{figure*}[!htbp]
    \includegraphics[width = 0.98\textwidth ]{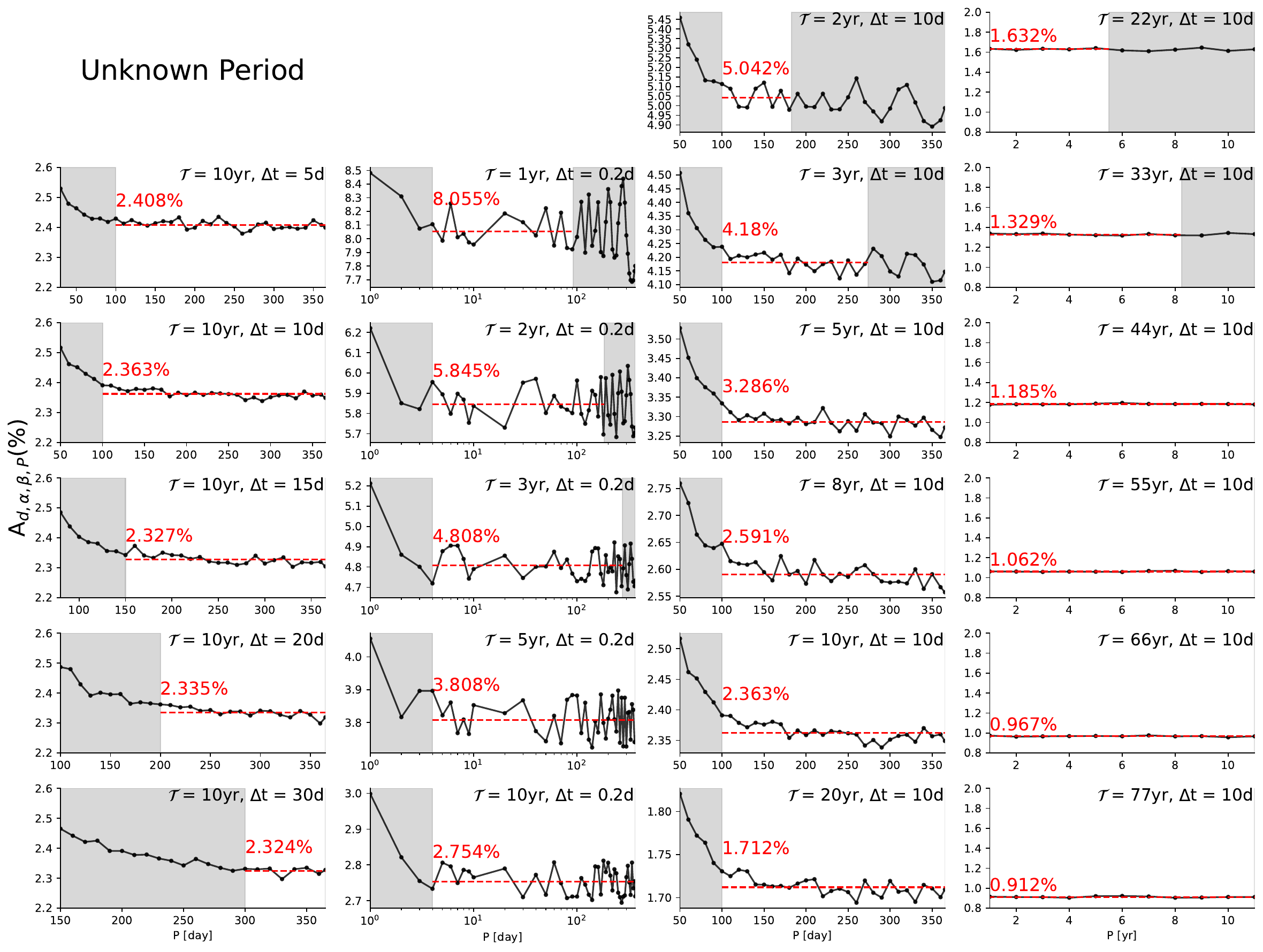}
    \caption{ The ``Unknown Period" scenario plot for 90\% detection ($\alpha$ = 0.1, $\beta$ = 0.1) 
 amplitude $A_{d,\alpha,\beta, P}$ as a function of $P$ for different ($\T$, $\Delta t$) and when \mbox{$\R_s = 100$ ton$^{-1}$ yr$^{-1}$}, \mbox{$\R_b = 0$ ton$^{-1}$ yr$^{-1}$} and \mbox{$\D = 100$ tons} in the simulation study. {\it 1$^{st}$ column:} \mbox{$\T = 10$ years}, $\Delta t$ varies from $5$ to $30$ days. {\it 2$^{nd}$ column:} \mbox{$\Delta t = 0.2$ days} and $\T$ varies from 1 to 10 years. {\it 3$^{rd}$ column:} \mbox{$\Delta t = 10$ days} and $\T$ varies from 2 to 20 years. {\it 4$^{th}$ column:} \mbox{$\Delta t = 10$ days} and $\T$ varies from 22 to 77 years. In each panel, the white region is where the amplitude $A_{d, \alpha,\beta, P}$ can be approximated by the average of the amplitudes over different frequencies $\overline{A}_{d, \alpha,\beta}$, and the red horizontal line is $\overline{A}_{d, \alpha,\beta}$. }
    \label{fig:Ps_Adcontour_UnknownP}
\end{figure*}

\begin{figure*}[!htbp]
    \includegraphics[width = 0.98\textwidth ]{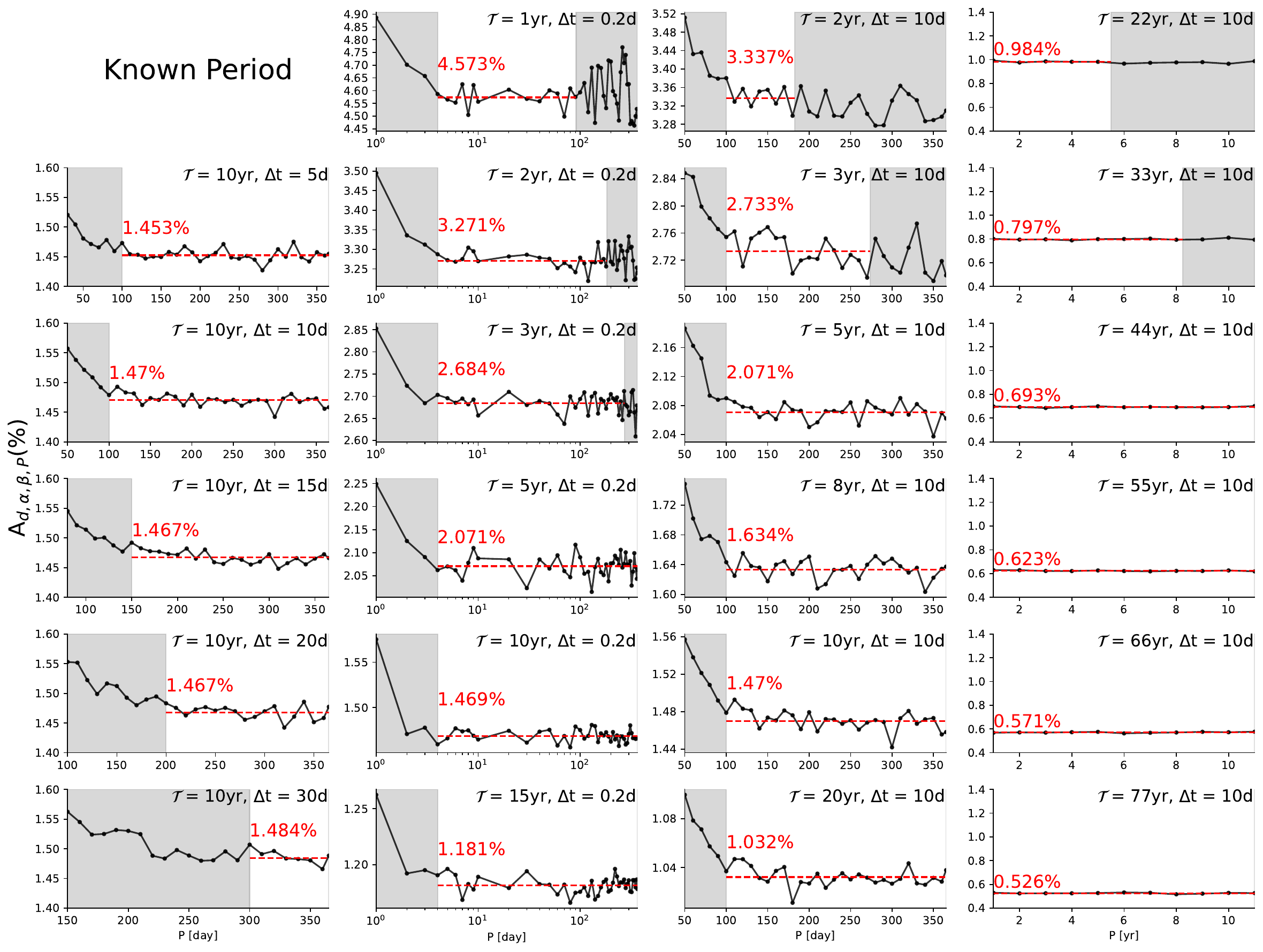}
    \caption{ Same as Figure~\ref{fig:Ps_Adcontour_UnknownP} but for known periods. }
    \label{fig:Ps_Adcontour_KnownP}
\end{figure*}

The simulation results showing how  $A_{d, \alpha, \beta}$ depends on different parameters as a function of $P$ are given in Figures~\ref{fig:Ps_Adcontour_UnknownP}~\ref{fig:Ps_Adcontour_KnownP}. Each panel is the $A_{d,\alpha = 0.1,\beta = 0.1, P}$ for a series of scanned frequencies ($P$) under a given $(\T, \Delta t)$. $A_{d,\alpha = 0.1,\beta = 0.1, P}$ is roughly constant when $\Delta t$ is small enough such that the reached $A_{d}$ is the same as the input $A_{d}$, $10 \Delta t<P<\T / 4$ when $\Delta t\geq 10$ days, and $20\Delta t<P<\T/ 4$ when $\Delta t < 10$ days. In addition, we found an interesting empirical relationship that by an equation \mbox{$y= ax^2 + bx+c$}, where \mbox{$y = \log ( \overline{A}_{d, \alpha,\beta}^2 \R_s \D \Delta t )$} and $x$ depends on the sample size of observation $N_t$ and the critical value. 
Specifically, $x = \log \{ \tau ( \M_{scan})/N_t \}$, where $\tau ( \M_{scan}) = max(1, \log(\M_{scan}))$. Under the ``Known Period scenario", $\M_{scan} = 1$ and $x = \log (1/N_t )$.
For a given $\alpha$, the critical value is constant under the ``Known Period scenario", and for ``Unknown Period scenario", with \mbox{$\M_{scan} \gg$ 1}, the critical value \mbox{$LS_{max, \alpha, 2} \approx CONST +\log(M_{scan})$}. The corresponding $x$ and the fitted $a$, $b$, $c$ for different parameters are showed in Table~\ref{tab: fit lnSNR}

\begin{table}[!htbp]
\caption{Fitted values of $a$, $b$, $c$ for the empirical relationship \mbox{$y= ax^2 + bx+c$} within [$x_{min}$, $x_{max}$] assuming \mbox{$\R_s$ = 100 ton$^{-1}$ yr$^{-1}$}, \mbox{$\R_b$ = 0}, \mbox{${\cal D} = 100$ ton}.
}
\begin{tabular}{cccccccccc}
\hline

& $x_{min}$ & $x_{max}$ & $\alpha$ & 
$\beta$ & $a$ & $b$ & $c$ \\

\hline
\multirow{2}{*}{\makecell{Known Period}} & \multirow{2}{*}{-10.218} & \multirow{2}{*}{-4.291} &0.1 & 0.1 & 3.004e-03 & 1.049 & 3.251 \\
& &&0.05 & 0.1 & 5.168e-03 & 1.077 & 3.529 \\
\hline

\multirow{2}{*}{\makecell{Unknown Period  
}} & \multirow{2}{*}{-7.5} & \multirow{2}{*}{-2.755} &0.1 & 0.1 & 2.344e-02 & 1.309 & 3.037 \\
& &&0.05 & 0.1 & 2.598e-02 & 1.345 & 3.224\\
\hline
\end{tabular}
\label{tab: fit lnSNR}
\end{table}

Using the above empirical relations, to summarize the procedure to find $A_{d, \alpha, \beta}$, first, we select the proper $\Delta t$ and $\T$ based on possible $P$ such that $A_{d, \alpha, \beta, P} \approx \overline{A}_{d, \alpha,\beta}$. Adjust ${\cal M}_{scan}$ such that $LS_{max, \alpha, 1} \approx LS_{max, \alpha, 2}$ to ensure $H_0$ is consistent with the model, and obtain $\M_{scan}$ and $x$, then obtain \mbox{$\overline{A}_{d, \alpha,\beta}$ = $\sqrt{ \frac{e^{ax^2+bx+c}}{\R_s \D \Delta t}}$} using parameters in Table~\ref{tab: fit lnSNR} and obtain 
\begin{equation}
    \overline{A}_{d, \alpha,\beta}^{'} =\overline{A}_{d, \alpha,\beta} \times \sqrt{\frac{\R_s}{\R_s^{'}}} \sqrt{\frac{\D}{\D^{'}}} \sqrt{1+\frac{\R_b^{'}}{{\cal R}_{s}^{'}}} 
    \label{eqn:scale_Ad}
\end{equation} via Eqn~\ref{eqn:1004}. Then, the nearby amplitudes are scanned for more accurate simulations. The detector size $\D^{'}$ can be obtained similarly via $\D^{'} = \D \times \left( \frac{\overline{A}_{d, \alpha,\beta} }{A_{ecc} }\right)^{2} \frac{\R_s}{\R_s^{'}} \left( 1+\frac{\R_b^{'}}{{\cal R}_{s}^{'}} \right)$.

\par Figure~\ref{fig:scaleVSyr_ES}~\ref{fig:scaleVSyr_8B}~\ref{fig:scaleVSyr_atmNu} are the application of this method to estimate $\overline{A}_{d, \alpha=0.1,\beta=0.1}^{'}$ as a function of runtime ${\cal T}$. We choose \mbox{$\Delta t$ = 0.05 days} for \mbox{$P$ = 1 day}, and \mbox{$\Delta t$ = 10 days} for \mbox{$P$ = 1 yr} and 11 yrs, such that $A_{d, \alpha, \beta, P}^{'} \approx \overline{A}_{d, \alpha,\beta}^{'}$ for these periods. The minimum runtime ${\cal T}$ when $\overline{A}_{d, \alpha=0.1,\beta=0.1}^{'}$ meets $A_{ecc} = 0.03342$ or $A_{d, DN, max} = 0.00891$ or $A_{atm}$ can be interpreted as the above physical amplitudes are the 90\% detectable amplitudes. Though the empirical results are fitting within [$x_{min}$, $x_{max}$], the results from simulating the nearby run time $\T$ is consistent with these estimated $\T$ when $x < x_{min}$ or $x > x_{max}$.

\section{Results \label{sec:results}}

\par We now move on to presenting the results for the time-varying analysis, focusing first on solar neutrinos. As above, we consider the electron neutrino scattering channel (ES) and nuclear recoil channels (CE$\nu$NS) separately, and the results are split into ``Known Period" and ``Unknown Period" scenarios. Starting with the ES channel, we take the end-point energy of $^{7}$Be and calculate R$_{\odot}$ in Equation~\ref{eq:eccentricity} by summing all the components of the solar neutrino flux (pp, CNO, $^{7}$Be, and pep) over a range \mbox{[1 keV, 650 keV]} for the ideal resolution + ideal efficiency case and \mbox{[0 keV, 790 keV]} for the smear+efficiency case, since the contribution outside of this range is dominated by 2$\nu\beta\beta$. The flux within these ranges is dominated by pp neutrinos. 

\par
\begin{figure}[!htbp]
    \includegraphics[width = 0.8\textwidth ]{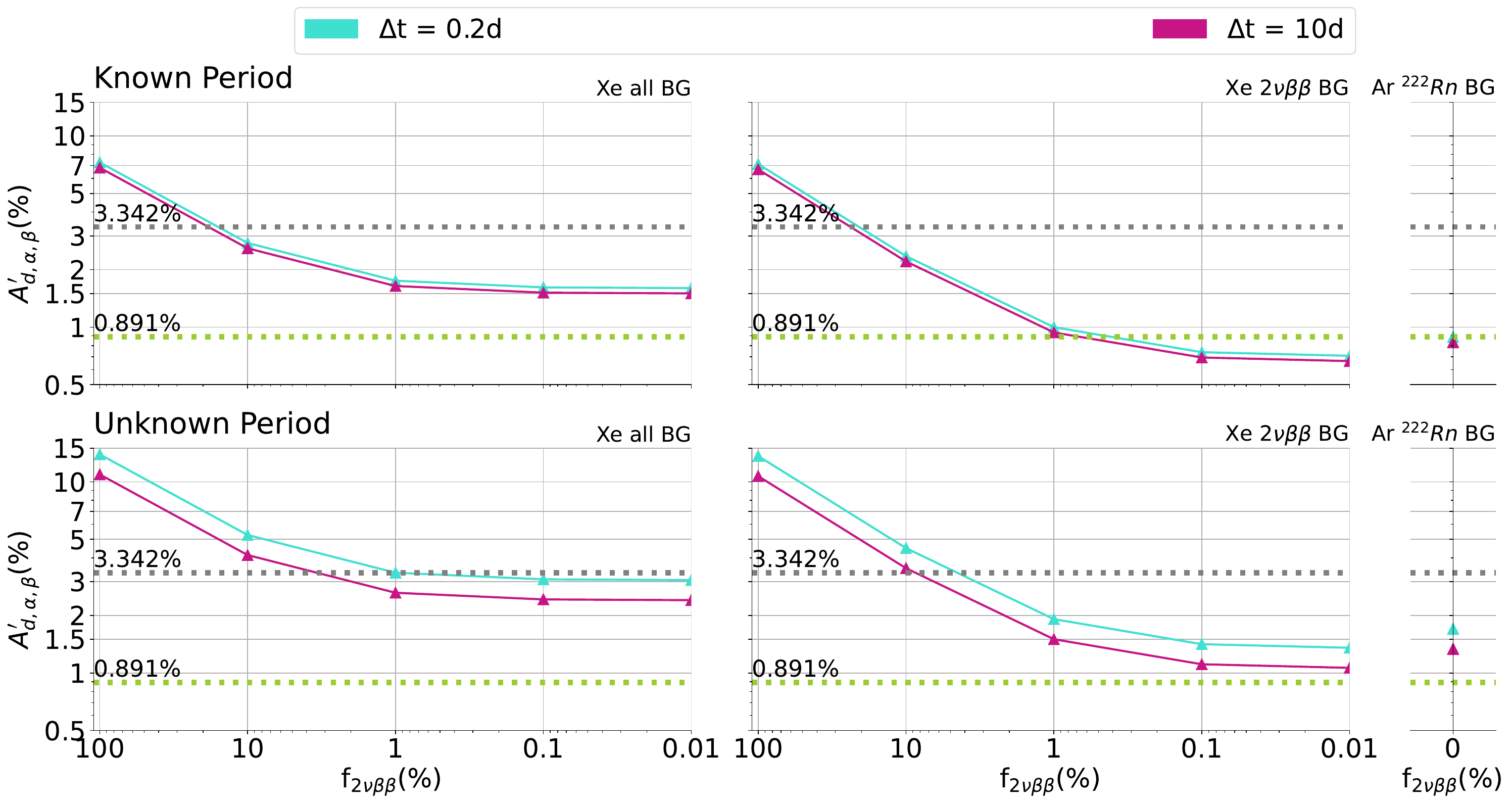}
    \caption{90\% detection ($\alpha$ = 0.1, $\beta$ = 0.1) amplitude $A_{d,\alpha,\beta}^{'}$ for solar components  through electron scattering obtained from Equation~\ref{eqn:1004} as a function of $2\nu \beta \beta$ fraction $f_{2\nu \beta \beta}$. Dashed lines are the amplitude of yearly modulation $A_{ecc} = 0.03342$ and the day-night modulation $A_{d, DN, max} = 0.00891$. The label ``all BG" includes $2\nu\beta\beta$, $^{222}$Rn, and $^{85}$Kr backgrounds. The assumed runtime is \mbox{$\T$ = 10 years}, and the detector size \mbox{$\D$ = 100 ton}. The time binnings and periods are $\Delta t= 0.2$ days for \mbox{$P$ = 1 day}, \mbox{$\Delta t= 10$ days} for \mbox{$P$ = 1 yr}. The curves show the spectrum including energy resolution and ``Xe100t-5" detector efficiency for Xenon (Figure~\ref{fig:detection_accep}), and 10 \% energy resolution with recoil energy threshold 100 keV only for Argon. Each curve is for a different time binning, $\Delta t$, as indicated. The top horizontal line shows the expected amplitude due to the eccentricity, and the bottom horizontal line shows the bound on the amplitude of the day-night modulation from Borexino.}
    \label{fig:ERamp_nubb}
\end{figure}

\begin{figure}[!htbp]
    \includegraphics[width = 0.8\textwidth ]{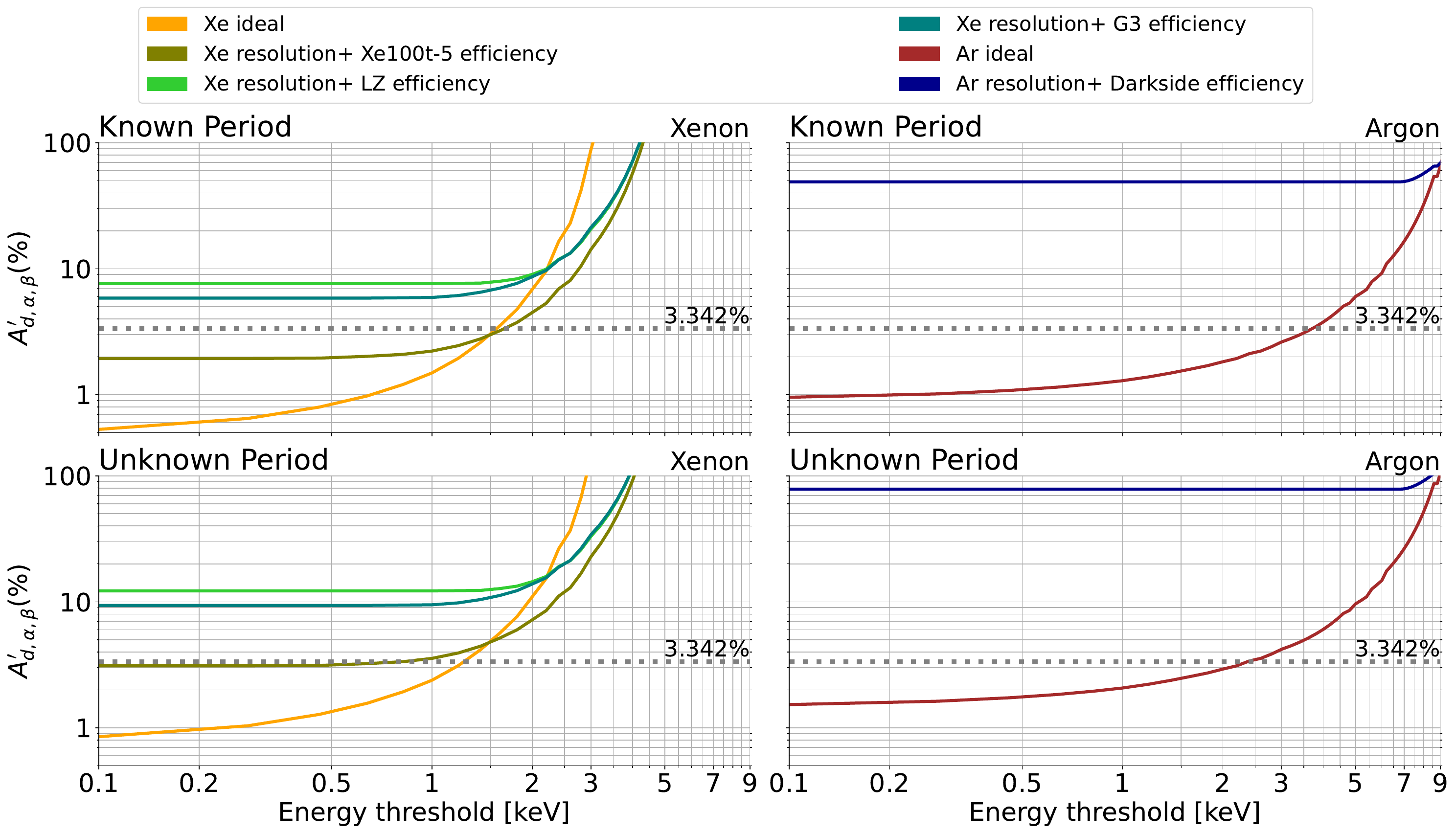}
    \caption{90\% detection ($\alpha$ = 0.1, $\beta$ = 0.1) amplitude $A_{d,\alpha,\beta}^{'}$ for $^{8}$B solar components through CE$\nu$NS  obtained from Equation~\ref{eqn:1004} as a function of recoil energy threshold, using \mbox{$\Delta t= 10$ days}. The assumed runtime is \mbox{$\T$ = 10 years}, and the detector size is \mbox{$\D$ = 100 tons}. Curves are shown for different detector and efficiency models as indicated. The dashed line is the amplitude of yearly modulation $A_{ecc} = 0.03342$.}
    \label{fig:NRamp_thrdeff}
\end{figure}

\begin{figure*}
    \centering
    \includegraphics[width = 0.98\textwidth ]{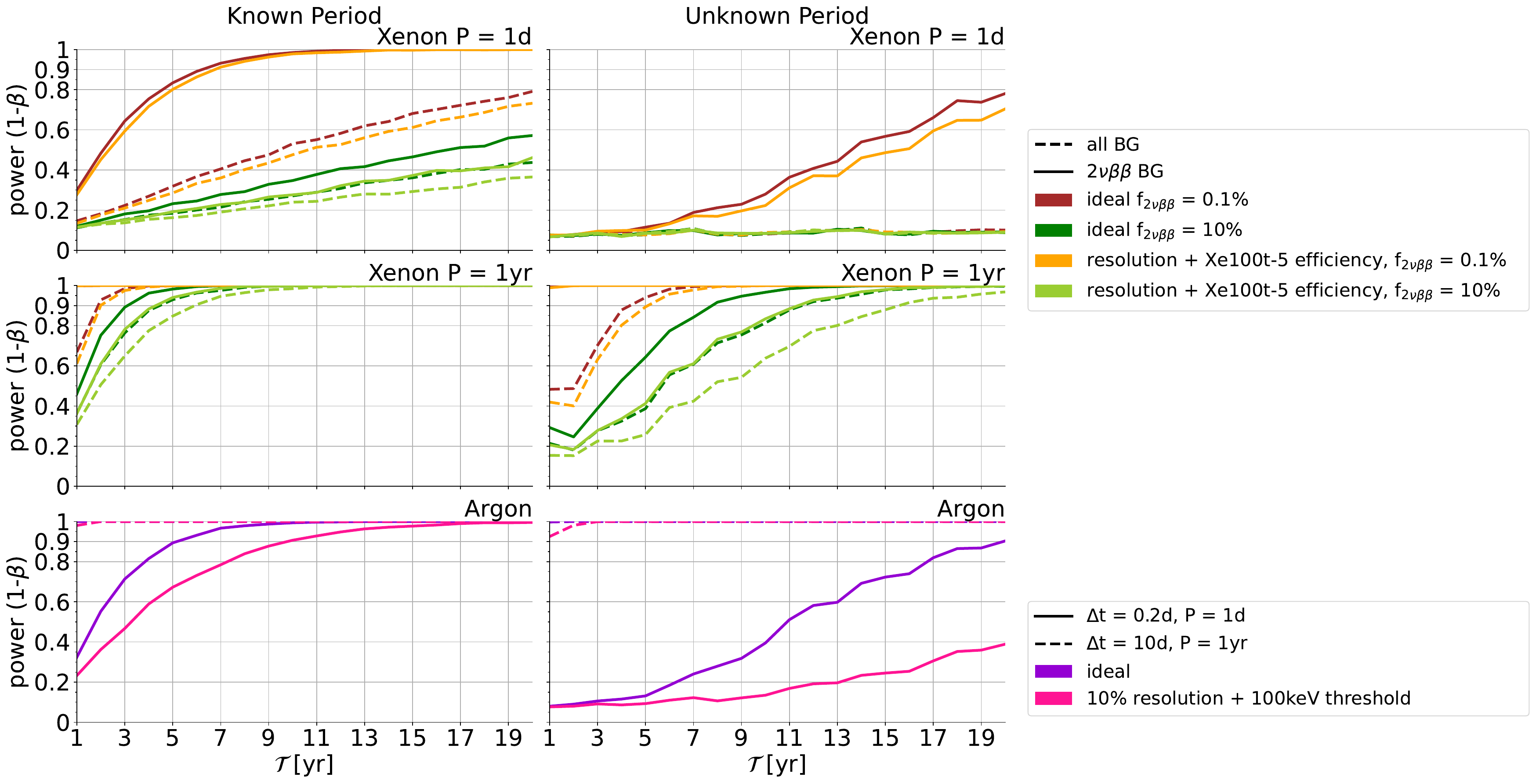}
    \caption{The power for detecting time variation of solar components (summing over pp, $^{7}$Be, CNO and pep components) through electron scattering, under the assumption of the ``Known Period" (left panel) and ``Unknown Period"  (right panel) scenarios described in the text. We take $\alpha = 0.1$, and plot the results as a function of detector runtime ${\cal T}$. The yearly modulation has an amplitude of $A_{ecc} = 0.03342$, due to the eccentricity of the Earth orbit, and the daily modulation has an amplitude of $A_{d, DN, max} = 0.00891$, which matches the Borexino upper limits on the daily modulation.  The assumed detector size is $\D$ = 100 ton. {\it Top row}: Xenon detection of $A_{d,DN, max}$ for a time binning of $\Delta t= 0.2$ days. {\it Middle row}: Xenon detection of $A_{ecc}$ with $\Delta t= 10$ days. {\it Bottom row}: Argon detection of $A_{ecc}$ with $\Delta t= 10$ days, and $A_{d, DN, max}$ with $\Delta t= 0.2$ days. For Xenon, ``all BG" includes $2\nu\beta\beta$, $^{222}$Rn and $^{85}$Kr. For Argon, the background includes $^{222}$Rn. }
    \label{fig:ECC_powerVSyr}
\end{figure*}

\begin{figure*}
    \centering
   \includegraphics[width = 0.85\textwidth ]{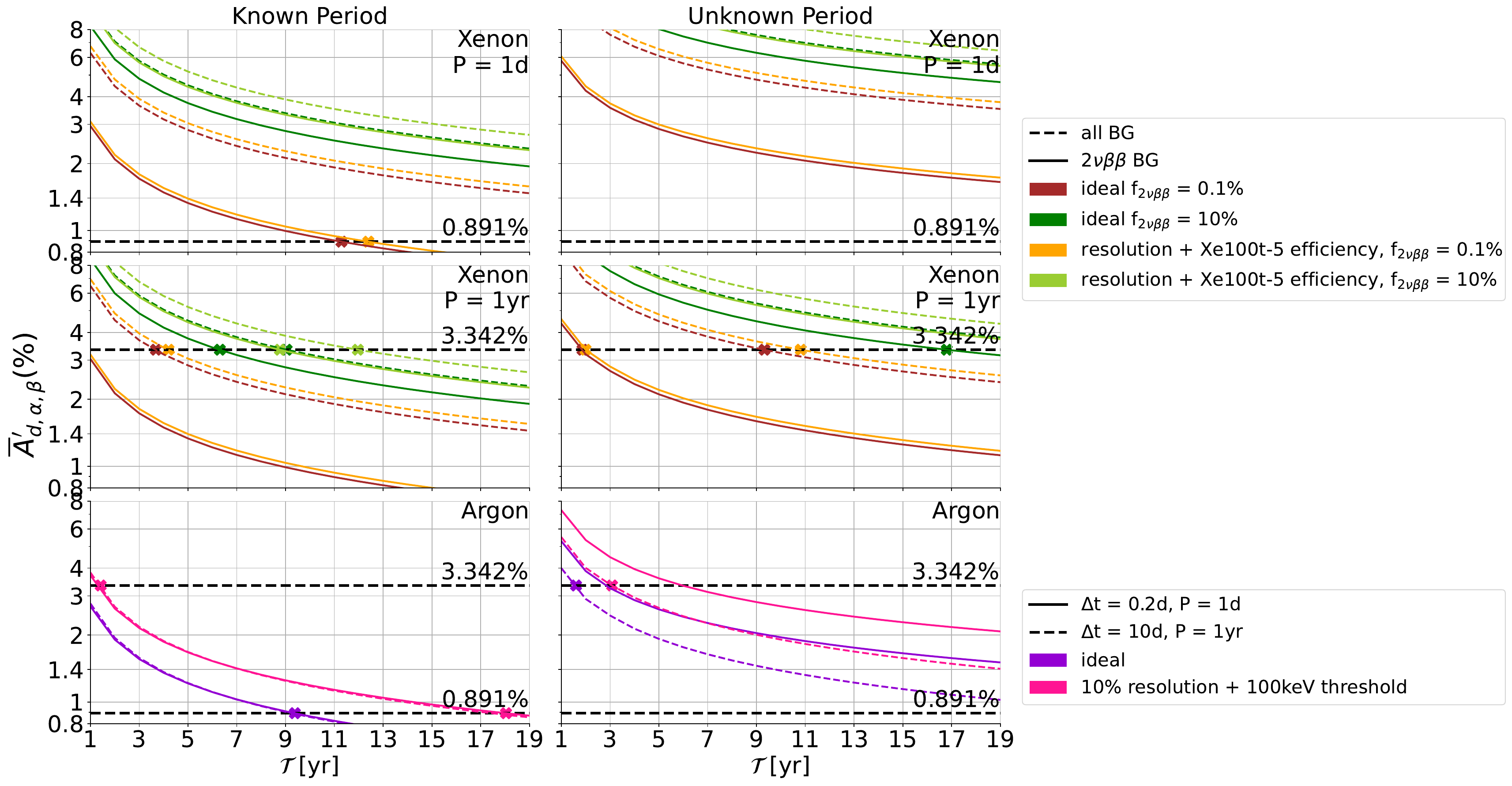}
    \caption{The detectable amplitude for solar components through electron scattering with 90\% power and 10\%  level of significance ($\alpha=0.1$, $\beta=0.1$), estimated from Equation~\ref{eqn:scale_Ad} as a function of detector runtime $\T$ for \mbox{$\D^{'} = 50$ tons}, under ``Known Period" and ``Unknown Period" scenarios. Dashed lines are the amplitude of yearly modulation $A_{ecc} = 0.03342$ and the day-night modulation $A_{d, DN, max} = 0.00891$. {\it Top row}: Xenon detection of $A_{d, DN, max}$ for a time binning of $\Delta t= 0.05$ days. {\it Middle row}: Xenon detection of $A_{ecc}$ with $\Delta t= 10$ days. {\it Bottom row}: Argon detection of $A_{ecc}$ with $\Delta t= 10$ days, and of $A_{d,DN, max}$ with $\Delta t= 0.05$ days. Points indicate the run time $\T$ when $\overline{A}^{'}_{d, \alpha=0.1, \beta=0.1}$ reaches $A_{ecc}$ or $A_{d,DN, max}$. For Xenon, ``all BG" includes $2\nu\beta\beta$, $^{222}$Rn and $^{85}$Kr. For Argon, the background includes $^{222}$Rn. }
    \label{fig:scaleVSyr_ES}
\end{figure*}

\begin{figure*}[!htbp]
    \centering
     \includegraphics[width = 0.9\textwidth ]{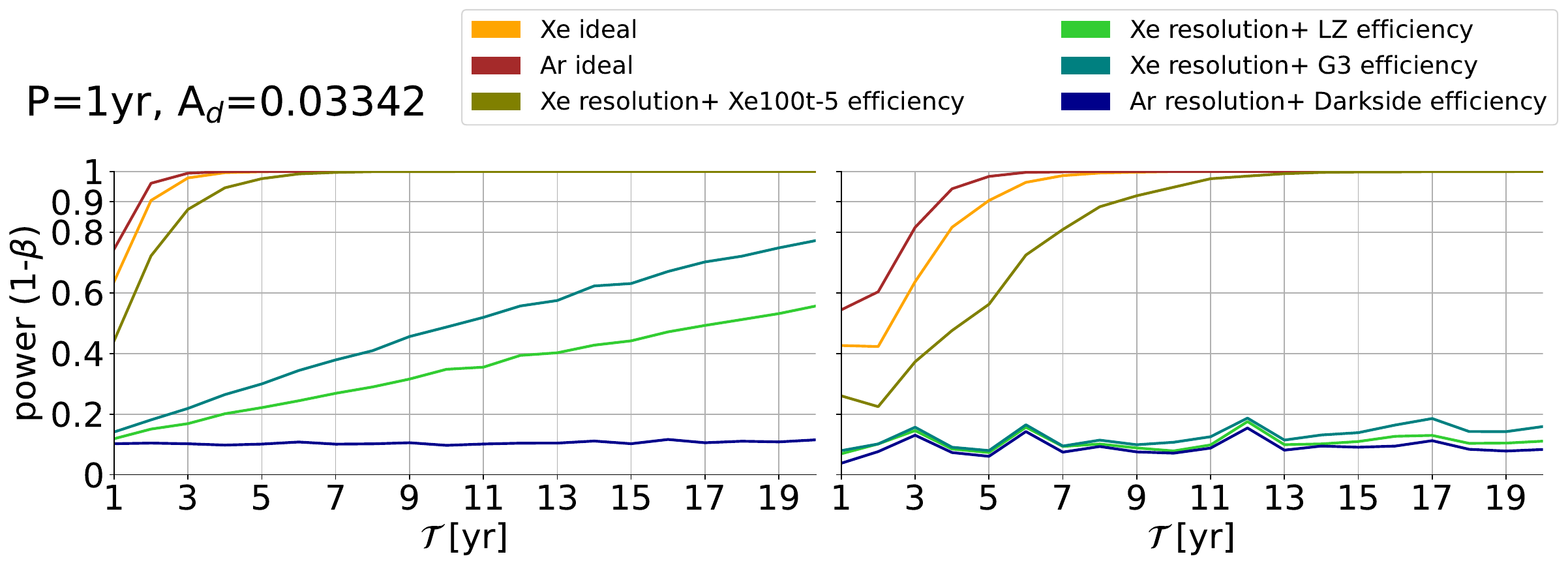}
    \caption{ The power of detecting time variation of $^{8}$B through CE$\nu$NS under ``Known Period" (left column) and ``Unknown Period"  (right column) scenarios with $\alpha = 0.1$, as a function of detector runtime ${\cal T}$. The assumed detector size is 100 ton. For Xenon assumptions are for an ideal detector and the ``Xe100t-5", ``LZ" and ``G3" detector efficiencies, and for Argon an ideal detector and the ``Darkside" detector efficiency.}
    \label{fig:NR_powerVSyr}
\end{figure*}

\begin{figure*}
    \includegraphics[width = 0.85\textwidth ]{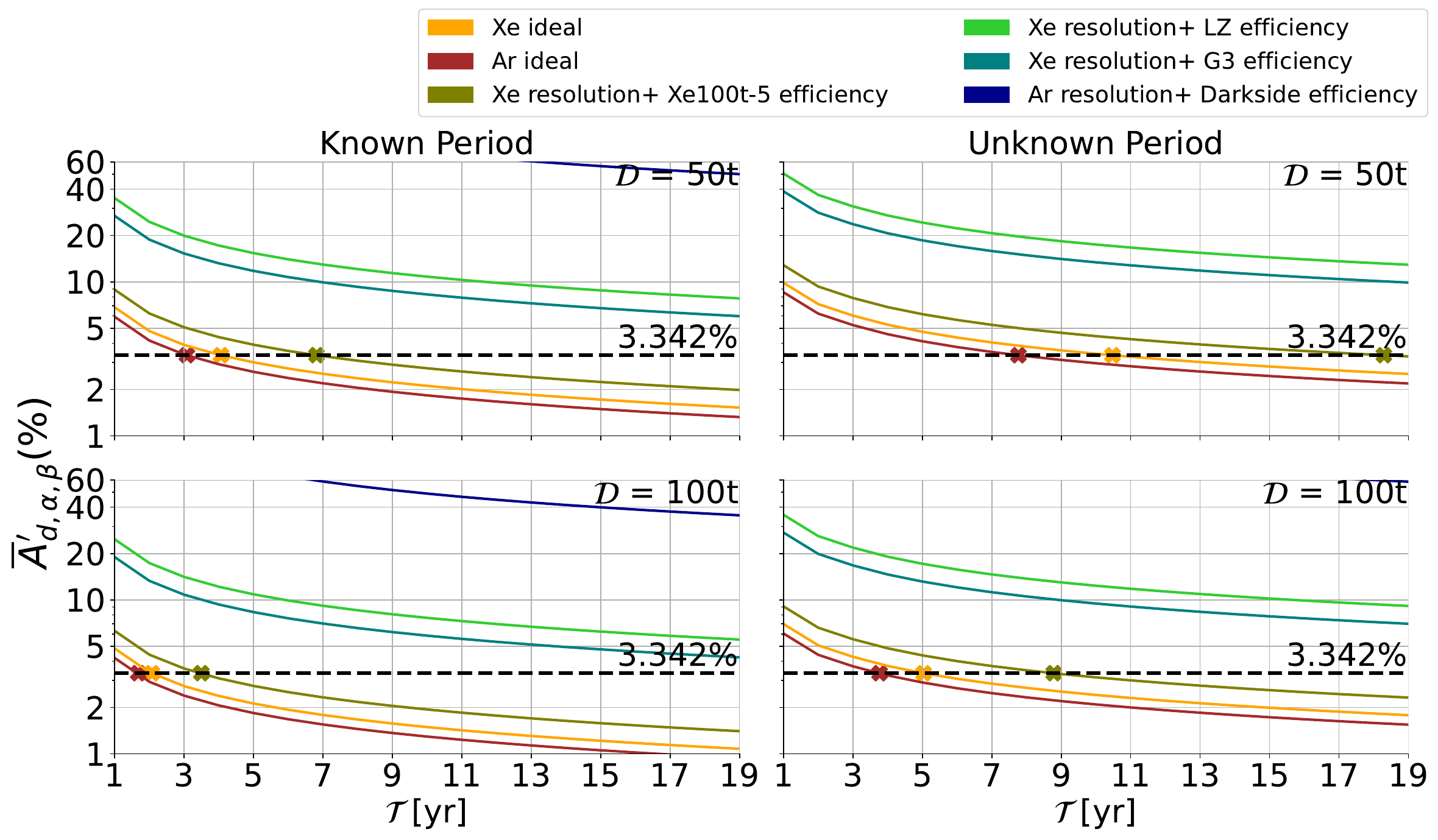}

    \caption{The detectable amplitude for $^{8}$B neutrinos through CE$\nu$NS with 90\% power and 10\%  level of significance ($\alpha=0.1$, $\beta=0.1$), estimated from Equation~\ref{eqn:scale_Ad} as a function of detector runtime $\T$, for \mbox{$\D^{'} = 50, 100$ tons}, under ``Known Period" and ``Unknown Period" scenarios, when \mbox{$\Delta t = 10$ days}. The dashed line is the amplitude of yearly modulation $A_{ecc} = 0.03342$. Crosses indicate the run time $\T$ when $\overline{A}^{'}_{d, \alpha=0.1, \beta=0.1}$ reaches $A_{ecc}$. Configurations of ($\R_s^{'}$, $\R_b^{'}$, $\D^{'}$) from different detector target, resolution, and efficiency are indicated.
 }
    \label{fig:scaleVSyr_8B}
\end{figure*}

\begin{figure*}[!htbp]
    \includegraphics[width = 0.9\textwidth ]{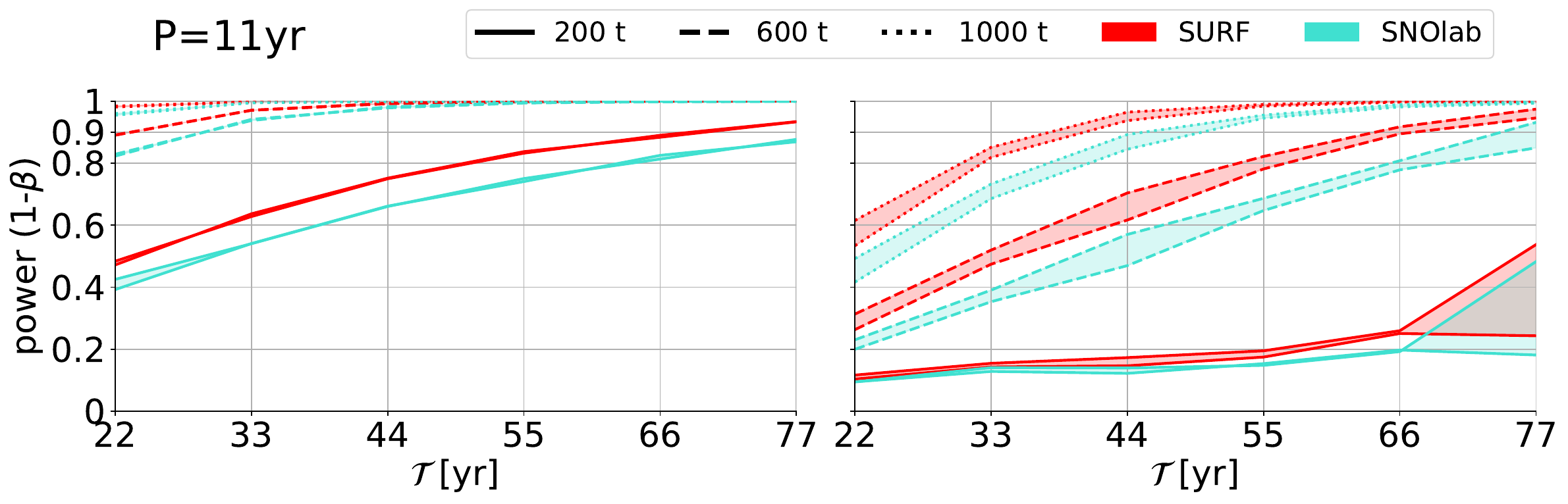}
    \caption{Power for detecting atmospheric neutrino time variation at SURF and SNOlab through nuclear recoils under ``Known Period" (left panel) and ``Unknown Period" (right panel) scenarios for $\alpha = 0.1$, as a function of detector runtime ${\cal T}$. The assumed detector sizes ${\cal D}$ are 200, 600, 1000 tons. Bands indicate the differences between \mbox{$\Delta t$ = $10$ days} and \mbox{$\Delta t$ = $30$ days} for assumptions of an ideal detector.}
    \label{fig:atmNu_loc_powerVSyr}
\end{figure*}

\begin{figure*}
    \centering
   \includegraphics[width = 0.8\textwidth ]{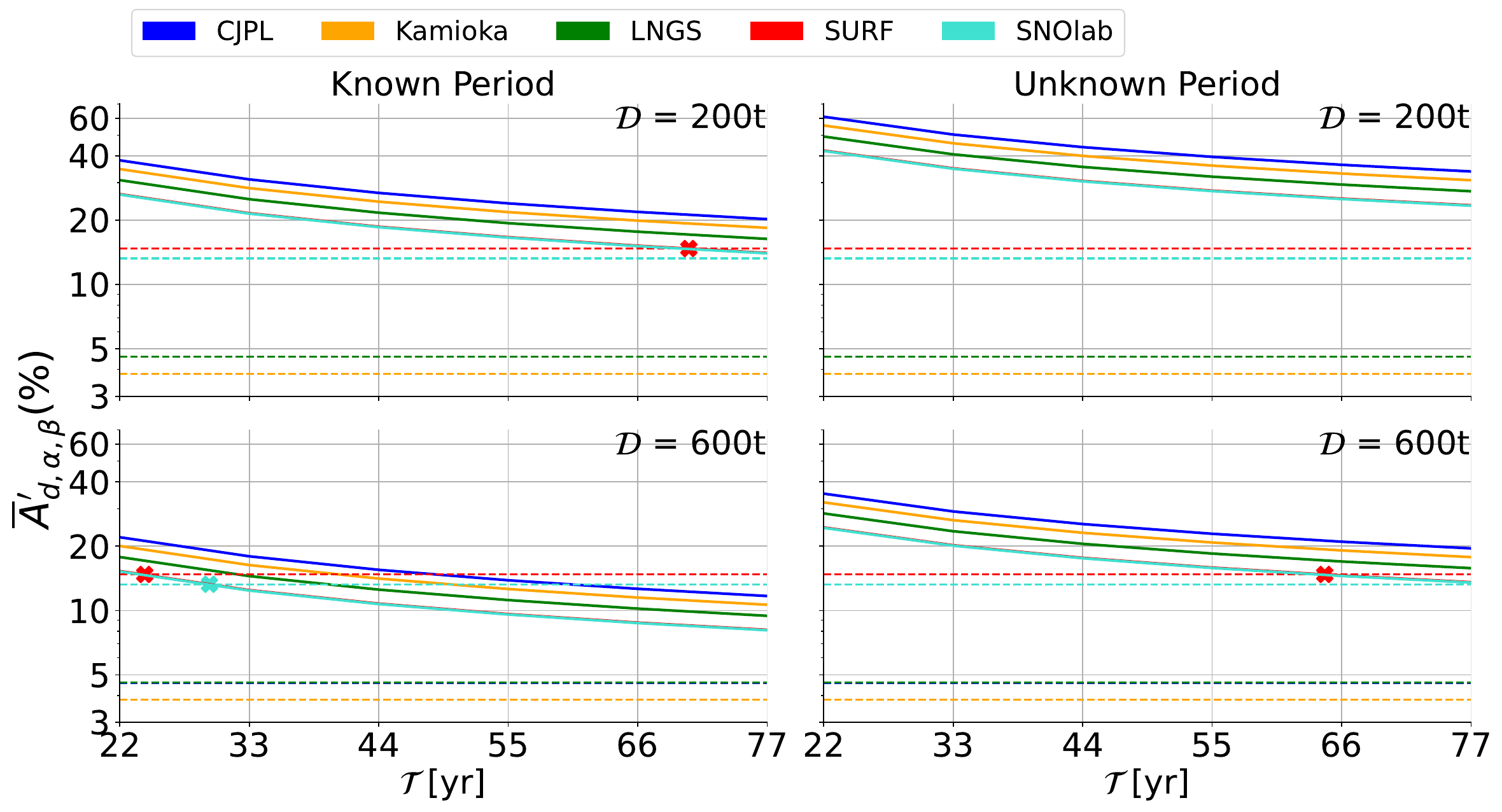}
    \caption{The detectable amplitude for atmospheric neutrinos through CE$\nu$NS with 90\% power and 10\%  level of significance ($\alpha=0.1$, $\beta=0.1$), estimated via Equation~\ref{eqn:scale_Ad} as a function of detector runtime $\T$, for \mbox{$\D^{'} = 200, 600$ tons}, under ``Known Period" and ``Unknown Period" scenarios, when \mbox{$\Delta t = 30$ days}. The dashed horizontal lines are $A_{atm}$ in Section~\ref{subsec:atm}, and each color is a different location as indicated. Crosses indicate the run time $\T$ when $\overline{A}^{'}_{d, \alpha=0.1, \beta=0.1}$ reaches $A_{atm}$ at the corresponding location. A xenon detector with ideal resolution and efficiency is assumed. }
    \label{fig:scaleVSyr_atmNu}
\end{figure*}

 We first show the results as a function of different depletion of backgrounds for ES, and as a function of different energy thresholds for CE$\nu$NS, under the fiducial case of a 100-ton detector, with a run time of 10 years and a fixed time-bin $\Delta t$ (Figure~\ref{fig:ERamp_nubb}~\ref{fig:NRamp_thrdeff}). Then we show the results as a function of detector run-time for a fixed time-bin $\Delta t$, detector size $\D$ for ES and CE$\nu$NS (Figures~\ref{fig:ECC_powerVSyr}~\ref{fig:scaleVSyr_ES}~\ref{fig:NR_powerVSyr}~\ref{fig:scaleVSyr_8B}~\ref{fig:atmNu_loc_powerVSyr}~\ref{fig:scaleVSyr_atmNu}) .

\par Figure~\ref{fig:ERamp_nubb} shows the required amplitude at which the probability of detecting the signal is $1-\beta = 0.9$, at $\alpha = 0.1$. Indicated on the figure as the upper horizontal line is the modulation amplitude induced by the eccentricity. We see that for a 2$\nu\beta\beta$ fraction $\lesssim 10\%$ at the assumed exposure the experiment will be sensitive to the time variation from eccentricity. Also shown as the lower horizontal lines are the upper limits on the day-night modulation of the ${}^7$Be component from Borexino, \mbox{A$_{d}$ = $A_{DN}/ \sqrt{2} \approx 0.009$}~\cite{Borexino:2022khe}. This shows that a 2$\nu\beta\beta$ fraction in Xenon at a level $\sim 1$\% has a  detectable amplitude with 90\% power similar to the best-fit amplitude A$_{d}$ from Borexino~\footnote{Although these two amplitudes are similar, they cannot be compared directly since the Borexino result comes from fitting the binned-data and our result comes from the likelihood ratio test.}. The Xenon measurements are sensitive to the lower energy pp component, for which there are no strong bounds on the time variation. 

\par Moving on to the CE$\nu$NS channel using the $^{8}$B flux, Figure~\ref{fig:NRamp_thrdeff} shows the required amplitude for the 90\% detection at $\alpha = 0.1$, as a function of nuclear recoil energy threshold. Shown are the results for different assumptions of the efficiency, in one case showing both perfect efficiency and in the remaining cases using the detector efficiency and energy resolution modeled from NEST simulations (Figure~\ref{fig:detection_accep}). From Figure~\ref{fig:NRamp_thrdeff} we see that the eccentricity is clearly detectable for the assumed exposure of 100 ton and 10 years for the ideal case. For the case with the efficiency and resolution corrections added, for thresholds $\lesssim 1$ keV there is a saturation and the eccentricity is nearly detectable given the specifications used. Note that since ${}^8$B neutrinos are detected via a neutral current interaction, there is no sensitivity to the day-night effect in the CE$\nu$NS channel (there could be sensitivity if radiative corrections were detectable~\cite{Mishra:2023jlq}). 

\par Figure~\ref{fig:ECC_powerVSyr} shows the power of detecting daily and yearly time variation through electron recoils as a function of detector runtime, for an assumed 100-ton detector size. We see that even for large $2\nu \beta \beta$ fractions, the eccentricity will be detectable at ``90\%" chance under the Known Period scenario with an exposure of $\lesssim 10$ years. Here we choose a time binning of $\Delta t= 10$ days, though we find that our results are relatively insensitive to the time binning. Similarly, under the Known Period scenario, the Borexino sensitivity can be achieved at ``90\%" chance for run times $\sim 7$ years when background includes only $2\nu\beta\beta$ and is depleted to 0.1\%.

\par Figure~\ref{fig:scaleVSyr_ES} shows the ``90\% detection" amplitude of detecting time variation through electron recoils as a function of detector runtime for $\Delta t$ = 0.05 days and an assumed 50-ton exposure. The 3\% eccentricity will be detectable with 90\% power under the Known Period scenario within $\sim 13$ years for large $2\nu \beta \beta$ fractions and other backgrounds, while $2\nu \beta \beta$ needs to be depleted down to $\lesssim$ 0.1\% to achieve the similar result under the ``Unknown Period scenario". For daily variations, a nearly-fully depleted detector will be able to match the Borexino upper limits on daily time variation for a run time of $\lesssim 13$ years under the Known Period scenario. 

\par~Figure~\ref{fig:NR_powerVSyr} shows the power of detecting time variation of $^8$B as a function of detector runtime, for an assumed detector size of 100 ton. Here we see that for an idealized detector, yearly time modulation is detectable at ``90\%" for a runtime of $\sim 3$ years under the Known Period scenario and $\sim 5$ years under the Unknown Period scenario. We note that this is similar to the time for detection of eccentricity using the electron recoil channel, for a low $2 \nu \beta \beta$ fraction. For the ``Xe100t-5" efficiency model, the detection timescale increases to $\lesssim$ 15 years. 

\par Figure~\ref{fig:scaleVSyr_8B} shows amplitude for 90\% detection of time variation of an arbitrary amplitude for $^8$B as a function of detector runtime for $\Delta t$ = 10 days and an assumed 50-ton or 100-ton exposure. From this we again see that the eccentricity will be detectable with 90\% power under the Known Period scenario within $\lesssim 10$ years for idealized detector or the resolution+``Xe100t-5" efficiency detector. Under the Unknown Period scenario, an idealized detector is needed to achieve the similar run time for a 50-ton or 100-ton detector. 

\par We finally checked the prospects for detection of time variation of atmospheric neutrinos. The resulting power for detecting the modulation of atmospheric neutrinos from Section~\ref{subsec:atm} at each location (the amplitude $A_{atm}$ is 0.0459 at CJPL, 0.0382 at Kamioka, 0.0461 at LNGS, 0.1475 at SURF and 0.1327 at SNOlab) is shown in Figure~\ref{fig:atmNu_loc_powerVSyr}, for the case of an ideal detector with no efficiency or energy resolution correction. Given that the signal is weaker than the signal for solar neutrinos, in this case we consider larger detector volumes of $200$, $600$, and $1000$ ton, and plot the power as a function of the experimental run time. Figure~\ref{fig:scaleVSyr_atmNu} shows the ``90\%" amplitude of detecting time variation of atmospheric neutrinos as a function of detector runtime for $\Delta t$ = 30 days and an assumed 200-ton or 600-ton exposure. Though the experimental volumes considered in Figures ~\ref{fig:atmNu_loc_powerVSyr}~\ref{fig:scaleVSyr_atmNu} are larger than that in our solar neutrino analysis, they provide a sense of what will be required to detect the atmospheric time variation.

\section{Conclusions \label{sec:conclusions}}

\par In this paper we have studied the prospects for detecting time variation of solar and atmospheric neutrinos in future large-scale Xenon and Argon dark matter detectors. We have developed rigorous statistical methods for detecting a time-varying signal, and compared the statistical methods under the assumptions that we are searching for a known periodicity or an unknown periodicity. 

\par For time varying signals, we have focused on detecting solar neutrinos through ES and CE$\nu$NS. In the ES channel, for a 100-ton Xenon detector running for 10 years with a Xenon-136 fraction of $\lesssim 0.1\%$, a time-variation amplitude of $\sim$ 1\% is detectable for the experimental configurations that we study. This is sufficient to detect time variation due to eccentricity. For this same run time, Xenon experiments will achieve similar sensitivity to the Borexino experiment for daily modulations that will be induced from the propagation of electron neutrinos through the Earth. Both of these results will hold if the $2\nu \beta \beta$ fraction can be reduced to $\lesssim 0.01$.

\par In the CE$\nu$NS channel, we find that for a threshold of 1 keV, a detector with ideal efficiency and energy resolution will be sensitive the a yearly variation in the flux. For a detector with efficiency and resolution similar to current Xenon experiments, detectors will approach sensitivity to the eccentricity for thresholds $\lesssim 1$ keV. In comparison to previous studies which considered the time variation in the nuclear recoil channel due to the eccentricity~\cite{2015Davis}, we add the more realistic detector scenarios to our modeling. The detector modeling effects most strongly affect the CE$\nu$NS channel, whereas the ES channel is most strongly affected by the $2\nu \beta \beta$ fraction, as discussed above. 

\par Dark matter experiments provide the unique possibility to measure the time variation of the solar neutrino flux through multiple detection channels. Detecting eccentricity in both channels would be important because it confirms the solar origin of the signal. Further, it provides a test for the long term stability of the detector, and provides a means to identify possible time-varying signals of unknown origin.

\section*{Acknowledgements} 
Y.~Z. and L.~E.~S. are supported by the DOE Grant No. DE-SC0010813.

\section{Appendix}\label{sec:appendix}

Table~\ref{tab: dt0.2} shows the critical values $LS_{max,\alpha,1}$ and $LS_{max,\alpha,2}$ for $\Delta t$ = 0.2 days and $\T$ from 1 yr to 20 yrs.

Table~\ref{tab:dt5days} shows the critical values $LS_{max,\alpha,1}$ and $LS_{max,\alpha,2}$ for $\Delta t$ = 5 days and $\T$ from 1 yr to 22 yrs.

Table~\ref{tab:dt10days} shows the critical values $LS_{max,\alpha,1}$ and $LS_{max,\alpha,2}$ for $\Delta t$ = 10 days and $\T$ from 1 yr to 22 yrs.

\begin{table}[!htbp]
\caption{Critical values for $\Delta t$ = 0.2 days, $f_{scan}$ = [$1/365.25 \, {\rm day}^{-1}$, 1 $\textrm{day}^{-1}$], $n_o$ = 10, $f_{adj}$=1.5}
\label{tab: dt0.2}
\begin{tabular}{l  rr|rr|rr|rr}
\hline
\multirow{2}{*}{$\T$ [yr]} & \multirow{2}{*}{$\M_{entire}$} & \multirow{2}{*}{$\M_{scan}$} &  \multicolumn{2}{c|}{$\alpha=0.1$} & \multicolumn{2}{c|}{$\alpha=0.05$} & \multicolumn{2}{c}{$\alpha=0.02$}\\

&&&$LS_{max,\alpha,1}$ & $LS_{max,\alpha,2}$ & $LS_{max,\alpha,1}$ & $LS_{max,\alpha,2}$ & $LS_{max,\alpha,1}$ & $LS_{max,\alpha,2}$ \\
\hline
1  & 12175   & 2429  & 10.05                   & 9.83                    & 10.77                    & 10.56                    & 11.7                     & 11.37                    \\
2  & 24350   & 4859  &                    10.74                   & 10.42                   & 11.46                    & 11.25                    & 12.39                    & 12.27                    \\
3  & 36525   & 7288  &                     11.14                   & 10.98                   & 11.86                    & 11.71                    & 12.8                     & 12.87                    \\
4  & 48700   & 9718  &                     11.43                   & 11.18                   & 12.15                    & 11.89                    & 13.08                    & 12.68                    \\
5  & 60875   & 12148 &  11.66                   & 11.39                   & 12.38                    & 12.15                    & 13.31                    & 13.07                    \\
6 & 73050   & 14577 & 11.84                   & 11.73                   & 12.56                    & 12.47                    & 13.49                    & 13.28                    \\
7  & 85225   & 17007 & 11.99                   & 12.05                   & 12.71                    & 12.84                    & 13.64                    & 13.84                    \\
8  & 97400   & 19437 &  12.13                   & 11.89                   & 12.85                    & 12.7                     & 13.78                    & 13.79                    \\
9  & 109575  & 21866 &  12.24                   & 12.01                   & 12.96                    & 12.76                    & 13.89                    & 13.58                    \\
10 & 121750  & 24296 &  12.35                   & 12.14                   & 13.07                    & 12.87                    & 14.0                     & 13.83                    \\
11 & 133925  & 26726 &  12.44                   & 12.34                   & 13.16                    & 13.11                    & 14.1                     & 14.0                     \\
12 & 146100  & 29155 & 12.53                   & 12.49                   & 13.25                    & 13.24                    & 14.18                    & 14.33                    \\
13 & 158275  & 31585 &  12.61                   & 12.62                   & 13.33                    & 13.32                    & 14.26                    & 14.24                    \\
14 & 170450  & 34015 &  12.68                   & 12.78                   & 13.4                     & 13.53                    & 14.34                    & 14.46                    \\
15 & 182625  & 36444 &  12.75                   & 12.54                   & 13.47                    & 13.23                    & 14.41                    & 14.2                     \\
16 & 194800  & 38874 & 12.82                   & 12.61                   & 13.54                    & 13.33                    & 14.47                    & 14.21                    \\
17 & 206975  & 41304 & 12.88                   & 12.66                   & 13.6                     & 13.37                    & 14.53                    & 14.22 \\
18 & 219150  & 43733 & 12.94                   & 12.77                   & 13.66                    & 13.51                    & 14.59                    & 14.3                     \\
19 & 231325  & 46163 & 12.99                   & 12.75                   & 13.71                    & 13.45                    & 14.64                    & 14.53                    \\
20 & 243500  & 48593 &  13.04                   & 12.97                   & 13.76                    & 13.69                    & 14.69                    & 14.65                   \\
\hline
\end{tabular}
\end{table}

\begin{table}[!htbp]
\caption{Critical values for $\Delta t$ = 5 days, $f_{scan}$ = [$1/365.25 \, {\rm day}^{-1}$, $1/5 \, {\rm day}^{-1}$], $n_o$ = 10}
\begin{tabular}{llrr|rr|rr|rr}
\hline
\multirow{2}{*}{$f_{adj}$} & \multirow{2}{*}{$\T$ [yr]} & \multirow{2}{*}{$\M_{entire}$} & \multirow{2}{*}{$\M_{scan}$} & \multicolumn{2}{c|}{$\alpha=0.1$} & \multicolumn{2}{c|}{$\alpha=0.05$} & \multicolumn{2}{c}{$\alpha=0.02$}\\

&&&&$LS_{max,\alpha,1}$ & $LS_{max,\alpha,2}$ & $LS_{max,\alpha,1}$ & $LS_{max,\alpha,2}$ & $LS_{max,\alpha,1}$ & $LS_{max,\alpha,2}$ \\
\hline

7      &1  & 104     & 104   &  6.9                     & 6.79                    & 7.61                     & 7.44                     & 8.55                     & 8.39                     \\
6  & 2  & 243     & 243   & 7.74                    & 7.68                    & 8.46                     & 8.43                     & 9.4                      & 9.53                     \\
6  & 3  & 365     & 365   &  8.15                    & 8.66                    & 8.87                     & 9.92                     & 9.8                      & 12.29                    \\
6  & 4  & 487     & 487   & 8.44                    & 8.55                    & 9.16                     & 9.32                     & 10.09                    & 10.66                    \\
6  & 5  & 608     & 608   & 8.66                    & 8.44                    & 9.38                     & 9.14                     & 10.31                    & 9.96                     \\
6  & 6  & 730     & 730   & 8.84                    & 9.31                    & 9.56                     & 10.56                    & 10.49                    & 12.83                    \\
6  & 7  & 852     & 852   &  9.0                     & 8.96                    & 9.72                     & 9.67                     & 10.65                    & 10.62                    \\
6  & 8  & 974     & 974   & 9.13                    & 9.14                    & 9.85                     & 9.94                     & 10.78                    & 10.92                    \\
6  & 9  & 1095    & 1095  & 9.25                    & 9.14                    & 9.97                     & 9.85                     & 10.9                     & 10.84                    \\
6  & 10 & 1217    & 1217  &9.35                    & 9.23                    & 10.07                    & 9.98                     & 11.01                    & 11.03                    \\
6  & 11 & 1339    & 1339  & 9.45                    & 9.48                    & 10.17                    & 10.3                     & 11.1                     & 11.34                    \\
6  & 12 & 1461    & 1461  & 9.54                    & 10.0                    & 10.26                    & 11.05                    & 11.19                    & 13.2                     \\
6  & 13 & 1582    & 1582  & 9.62                    & 9.58                    & 10.34                    & 10.4                     & 11.27                    & 11.26                    \\
6  & 14 & 1704    & 1704  &  9.69                    & 9.71                    & 10.41                    & 10.52                    & 11.34                    & 11.64                    \\
6  & 15 & 1826    & 1826  &  9.76                    & 9.83                    & 10.48                    & 10.59                    & 11.41                    & 11.56                    \\
6  & 16 & 1948    & 1948  & 9.82                    & 9.94                    & 10.54                    & 10.68                    & 11.48                    & 11.63                    \\
6  & 17 & 2069    & 2069  & 9.89                    & 10.12                   & 10.61                    & 10.9                     & 11.54                    & 11.98                    \\
6  & 18 & 2191    & 2191  & 9.94                    & 9.8                     & 10.66                    & 10.59                    & 11.59                    & 11.51                    \\
6  & 19 & 2313    & 2313  &  10.0                    & 9.91                    & 10.72                    & 10.63                    & 11.65                    & 11.57                    \\
6  & 20 & 2435    & 2435  & 10.05                   & 10.03                   & 10.77                    & 10.77                    & 11.7                     & 11.68                    \\
6  & 21 & 2556    & 2556  & 10.1                    & 10.16                   & 10.82                    & 10.87                    & 11.75                    & 12.0                     \\
6  & 22 & 2678    & 2678  & 10.14                   & 10.12                   & 10.86                    & 10.83                    & 11.79                    & 11.9\\
\hline
\end{tabular}
\label{tab:dt5days}
\end{table}

\begin{table}[!htbp]
\caption{Critical values for $\Delta t$ = 10 days, $f_{scan}$ = [$1/365.25 \, {\rm day}^{-1}$, $1/10 \, {\rm day}^{-1}$], $n_o$ = 10}
\begin{tabular}{ll rr|rr|rr|rr}
\hline
\multirow{2}{*}{$f_{adj}$} & \multirow{2}{*}{$\T$ [yr]} & \multirow{2}{*}{$\M_{entire}$} & \multirow{2}{*}{$\M_{scan}$} & \multicolumn{2}{c|}{$\alpha=0.1$} & \multicolumn{2}{c|}{$\alpha=0.05$} & \multicolumn{2}{c}{$\alpha=0.02$}\\

&&&&$LS_{max,\alpha,1}$ & $LS_{max,\alpha,2}$ & $LS_{max,\alpha,1}$ & $LS_{max,\alpha,2}$ & $LS_{max,\alpha,1}$ & $LS_{max,\alpha,2}$ \\
\hline
7      &1  & 52      & 52    &  6.2                     & 5.76                    & 6.92                     & 6.42                     & 7.85                     & 7.28                     \\
 7      & 2  & 104     & 104   & 6.9                     & 6.88                    & 7.61                     & 7.62                     & 8.55                     & 8.63                     \\
6    & 3  & 182     & 182   & 7.45                    & 8.47                    & 8.17                     & 10.26                    & 9.11                     & 13.55                    \\
6    & 4  & 243     & 243   &  7.74                    & 7.57                    & 8.46                     & 8.35                     & 9.4                      & 9.42                     \\
6    & 5  & 304     & 304   & 7.97                    & 7.66                    & 8.69                     & 8.32                     & 9.62                     & 9.35                     \\
6    & 6 & 365     & 365   & 8.15                    & 9.07                    & 8.87                     & 10.64                    & 9.8                      & 14.11                    \\
6    & 7 & 426     & 426   & 8.3                     & 8.1                     & 9.02                     & 8.83                     & 9.96                     & 9.66                     \\
6    & 8 & 487     & 487   &  8.44                    & 8.42                    & 9.16                     & 9.21                     & 10.09                    & 10.35                    \\
6    & 9 & 547     & 547   & 8.55                    & 8.35                    & 9.27                     & 9.03                     & 10.21                    & 9.91                     \\
6    & 10 & 608     & 608   &  8.66                    & 8.4                     & 9.38                     & 9.2                      & 10.31                    & 10.11                    \\
6    & 11 & 669     & 669   & 8.76                    & 8.68                    & 9.48                     & 9.45                     & 10.41                    & 10.61                    \\
6    & 12 & 730     & 730   & 8.84                    & 9.92                    & 9.56                     & 12.1                     & 10.49                    & 16.08                    \\
6    & 13 & 791     & 791   & 8.92                    & 8.86                    & 9.64                     & 9.55                     & 10.58                    & 10.43                    \\
6    & 14 & 852     & 852   & 9.0                     & 8.86                    & 9.72                     & 9.55                     & 10.65                    & 10.64                    \\
6    & 15 & 913     & 913   &  9.07                    & 8.96                    & 9.79                     & 9.67                     & 10.72                    & 10.69                    \\
6    & 16 & 974     & 974   & 9.13                    & 9.23                    & 9.85                     & 10.06                    & 10.78                    & 11.24                    \\
6    & 17 & 1034    & 1034  &  9.19                    & 9.38                    & 9.91                     & 10.21                    & 10.84                    & 11.11                    \\
6    & 18 & 1095    & 1095  & 9.25                    & 9.12                    & 9.97                     & 9.88                     & 10.9                     & 10.79                    \\
6    & 19 & 1156    & 1156  & 9.3                     & 9.2                     & 10.02                    & 10.0                     & 10.95                    & 10.89                    \\
6    & 20 & 1217    & 1217  &  9.35                    & 9.33                    & 10.07                    & 10.03                    & 11.01                    & 11.04                    \\
6    & 21 & 1278    & 1278  & 9.4                     & 9.28                    & 10.12                    & 10.04                    & 11.05                    & 11.07                    \\
6    & 22 & 1339    & 1339  &   9.45                    & 9.45                    & 10.17                    & 10.23                    & 11.1                     & 11.23                   \\
\hline
\end{tabular}
\label{tab:dt10days}
\end{table}

\bibliography{apssamp}

\providecommand{\noopsort}[1]{}\providecommand{\singleletter}[1]{#1}%
\begin{thebibliography}{44}
\expandafter\ifx\csname natexlab\endcsname\relax\def\natexlab#1{#1}\fi
\expandafter\ifx\csname bibnamefont\endcsname\relax
  \def\bibnamefont#1{#1}\fi
\expandafter\ifx\csname bibfnamefont\endcsname\relax
  \def\bibfnamefont#1{#1}\fi
\expandafter\ifx\csname citenamefont\endcsname\relax
  \def\citenamefont#1{#1}\fi
\expandafter\ifx\csname url\endcsname\relax
  \def\url#1{\texttt{#1}}\fi
\expandafter\ifx\csname urlprefix\endcsname\relax\def\urlprefix{URL }\fi
\providecommand{\bibinfo}[2]{#2}
\providecommand{\eprint}[2][]{\url{#2}}

\bibitem[{\citenamefont{Aprile et~al.}(2018)}]{XENON:2018voc}
\bibinfo{author}{\bibfnamefont{E.}~\bibnamefont{Aprile}} \bibnamefont{et~al.}
  (\bibinfo{collaboration}{XENON}), \bibinfo{journal}{Phys. Rev. Lett.}
  \textbf{\bibinfo{volume}{121}}, \bibinfo{pages}{111302}
  (\bibinfo{year}{2018}), \eprint{1805.12562}.

\bibitem[{\citenamefont{Aalbers
  et~al.}(2022{\natexlab{a}})}]{LUX-ZEPLIN:2022qhg}
\bibinfo{author}{\bibfnamefont{J.}~\bibnamefont{Aalbers}} \bibnamefont{et~al.}
  (\bibinfo{collaboration}{LUX-ZEPLIN}) (\bibinfo{year}{2022}{\natexlab{a}}),
  \eprint{2207.03764}.

\bibitem[{\citenamefont{Aalbers et~al.}(2022{\natexlab{b}})}]{Aalbers:2022dzr}
\bibinfo{author}{\bibfnamefont{J.}~\bibnamefont{Aalbers}} \bibnamefont{et~al.}
  (\bibinfo{year}{2022}{\natexlab{b}}), \eprint{2203.02309}.

\bibitem[{\citenamefont{Billard et~al.}(2014)\citenamefont{Billard, Strigari,
  and Figueroa-Feliciano}}]{Billard:2013qya}
\bibinfo{author}{\bibfnamefont{J.}~\bibnamefont{Billard}},
  \bibinfo{author}{\bibfnamefont{L.}~\bibnamefont{Strigari}}, \bibnamefont{and}
  \bibinfo{author}{\bibfnamefont{E.}~\bibnamefont{Figueroa-Feliciano}},
  \bibinfo{journal}{Phys. Rev. D} \textbf{\bibinfo{volume}{89}},
  \bibinfo{pages}{023524} (\bibinfo{year}{2014}), \eprint{1307.5458}.

\bibitem[{\citenamefont{Dutta and Strigari}(2019)}]{Dutta:2019oaj}
\bibinfo{author}{\bibfnamefont{B.}~\bibnamefont{Dutta}} \bibnamefont{and}
  \bibinfo{author}{\bibfnamefont{L.~E.} \bibnamefont{Strigari}},
  \bibinfo{journal}{Ann. Rev. Nucl. Part. Sci.} \textbf{\bibinfo{volume}{69}},
  \bibinfo{pages}{137} (\bibinfo{year}{2019}), \eprint{1901.08876}.

\bibitem[{\citenamefont{Dent et~al.}(2016)\citenamefont{Dent, Dutta, Newstead,
  and Strigari}}]{Dent:2016iht}
\bibinfo{author}{\bibfnamefont{J.~B.} \bibnamefont{Dent}},
  \bibinfo{author}{\bibfnamefont{B.}~\bibnamefont{Dutta}},
  \bibinfo{author}{\bibfnamefont{J.~L.} \bibnamefont{Newstead}},
  \bibnamefont{and} \bibinfo{author}{\bibfnamefont{L.~E.}
  \bibnamefont{Strigari}}, \bibinfo{journal}{Phys. Rev. D}
  \textbf{\bibinfo{volume}{93}}, \bibinfo{pages}{075018}
  (\bibinfo{year}{2016}), \eprint{1602.05300}.

\bibitem[{\citenamefont{Dent et~al.}(2017)\citenamefont{Dent, Dutta, Newstead,
  and Strigari}}]{Dent:2016wor}
\bibinfo{author}{\bibfnamefont{J.~B.} \bibnamefont{Dent}},
  \bibinfo{author}{\bibfnamefont{B.}~\bibnamefont{Dutta}},
  \bibinfo{author}{\bibfnamefont{J.~L.} \bibnamefont{Newstead}},
  \bibnamefont{and} \bibinfo{author}{\bibfnamefont{L.~E.}
  \bibnamefont{Strigari}}, \bibinfo{journal}{Phys. Rev. D}
  \textbf{\bibinfo{volume}{95}}, \bibinfo{pages}{051701}
  (\bibinfo{year}{2017}), \eprint{1607.01468}.

\bibitem[{\citenamefont{O'Hare et~al.}(2015)\citenamefont{O'Hare, Green,
  Billard, Figueroa-Feliciano, and Strigari}}]{OHare:2015utx}
\bibinfo{author}{\bibfnamefont{C.~A.~J.} \bibnamefont{O'Hare}},
  \bibinfo{author}{\bibfnamefont{A.~M.} \bibnamefont{Green}},
  \bibinfo{author}{\bibfnamefont{J.}~\bibnamefont{Billard}},
  \bibinfo{author}{\bibfnamefont{E.}~\bibnamefont{Figueroa-Feliciano}},
  \bibnamefont{and} \bibinfo{author}{\bibfnamefont{L.~E.}
  \bibnamefont{Strigari}}, \bibinfo{journal}{Phys. Rev. D}
  \textbf{\bibinfo{volume}{92}}, \bibinfo{pages}{063518}
  (\bibinfo{year}{2015}), \eprint{1505.08061}.

\bibitem[{\citenamefont{Davis}(2015)}]{2015Davis}
\bibinfo{author}{\bibfnamefont{J.~H.} \bibnamefont{Davis}},
  \bibinfo{journal}{JCAP} \textbf{\bibinfo{volume}{03}}, \bibinfo{pages}{012}
  (\bibinfo{year}{2015}), \eprint{1412.1475}.

\bibitem[{\citenamefont{{Sassi} et~al.}(2021)\citenamefont{{Sassi},
  {Dinmohammadi}, {Heikinheimo}, {Mirabolfathi}, {Nordlund}, {Safari}, and
  {Tuominen}}}]{2021Sassi}
\bibinfo{author}{\bibfnamefont{S.}~\bibnamefont{{Sassi}}},
  \bibinfo{author}{\bibfnamefont{A.}~\bibnamefont{{Dinmohammadi}}},
  \bibinfo{author}{\bibfnamefont{M.}~\bibnamefont{{Heikinheimo}}},
  \bibinfo{author}{\bibfnamefont{N.}~\bibnamefont{{Mirabolfathi}}},
  \bibinfo{author}{\bibfnamefont{K.}~\bibnamefont{{Nordlund}}},
  \bibinfo{author}{\bibfnamefont{H.}~\bibnamefont{{Safari}}}, \bibnamefont{and}
  \bibinfo{author}{\bibfnamefont{K.}~\bibnamefont{{Tuominen}}},
  \bibinfo{journal}{\prd} \textbf{\bibinfo{volume}{104}}, \bibinfo{eid}{063037}
  (\bibinfo{year}{2021}), \eprint{2103.08511}.

\bibitem[{\citenamefont{Cerde\~no et~al.}(2016)\citenamefont{Cerde\~no,
  Fairbairn, Jubb, Machado, Vincent, and B\oe{}hm}}]{Cerdeno:2016sfi}
\bibinfo{author}{\bibfnamefont{D.~G.} \bibnamefont{Cerde\~no}},
  \bibinfo{author}{\bibfnamefont{M.}~\bibnamefont{Fairbairn}},
  \bibinfo{author}{\bibfnamefont{T.}~\bibnamefont{Jubb}},
  \bibinfo{author}{\bibfnamefont{P.~A.~N.} \bibnamefont{Machado}},
  \bibinfo{author}{\bibfnamefont{A.~C.} \bibnamefont{Vincent}},
  \bibnamefont{and} \bibinfo{author}{\bibfnamefont{C.}~\bibnamefont{B\oe{}hm}},
  \bibinfo{journal}{JHEP} \textbf{\bibinfo{volume}{05}}, \bibinfo{pages}{118}
  (\bibinfo{year}{2016}), \bibinfo{note}{[Erratum: JHEP 09, 048 (2016)]},
  \eprint{1604.01025}.

\bibitem[{\citenamefont{Aristizabal~Sierra
  et~al.}(2019)\citenamefont{Aristizabal~Sierra, Dutta, Liao, and
  Strigari}}]{AristizabalSierra:2019ykk}
\bibinfo{author}{\bibfnamefont{D.}~\bibnamefont{Aristizabal~Sierra}},
  \bibinfo{author}{\bibfnamefont{B.}~\bibnamefont{Dutta}},
  \bibinfo{author}{\bibfnamefont{S.}~\bibnamefont{Liao}}, \bibnamefont{and}
  \bibinfo{author}{\bibfnamefont{L.~E.} \bibnamefont{Strigari}},
  \bibinfo{journal}{JHEP} \textbf{\bibinfo{volume}{12}}, \bibinfo{pages}{124}
  (\bibinfo{year}{2019}), \eprint{1910.12437}.

\bibitem[{\citenamefont{Aristizabal~Sierra
  et~al.}(2022)\citenamefont{Aristizabal~Sierra, De~Romeri, Flores, and
  Papoulias}}]{AristizabalSierra:2021kht}
\bibinfo{author}{\bibfnamefont{D.}~\bibnamefont{Aristizabal~Sierra}},
  \bibinfo{author}{\bibfnamefont{V.}~\bibnamefont{De~Romeri}},
  \bibinfo{author}{\bibfnamefont{L.~J.} \bibnamefont{Flores}},
  \bibnamefont{and} \bibinfo{author}{\bibfnamefont{D.~K.}
  \bibnamefont{Papoulias}}, \bibinfo{journal}{JCAP}
  \textbf{\bibinfo{volume}{01}}, \bibinfo{pages}{055} (\bibinfo{year}{2022}),
  \eprint{2109.03247}.

\bibitem[{\citenamefont{Zhuang et~al.}(2022)\citenamefont{Zhuang, Strigari, and
  Lang}}]{Zhuang:2021rsg}
\bibinfo{author}{\bibfnamefont{Y.}~\bibnamefont{Zhuang}},
  \bibinfo{author}{\bibfnamefont{L.~E.} \bibnamefont{Strigari}},
  \bibnamefont{and} \bibinfo{author}{\bibfnamefont{R.~F.} \bibnamefont{Lang}},
  \bibinfo{journal}{Phys. Rev. D} \textbf{\bibinfo{volume}{105}},
  \bibinfo{pages}{043001} (\bibinfo{year}{2022}), \eprint{2110.14723}.

\bibitem[{\citenamefont{O'Hare}(2021)}]{OHare:2021utq}
\bibinfo{author}{\bibfnamefont{C.~A.~J.} \bibnamefont{O'Hare}},
  \bibinfo{journal}{Phys. Rev. Lett.} \textbf{\bibinfo{volume}{127}},
  \bibinfo{pages}{251802} (\bibinfo{year}{2021}), \eprint{2109.03116}.

\bibitem[{\citenamefont{Appel et~al.}(2022)}]{Borexino:2022khe}
\bibinfo{author}{\bibfnamefont{S.}~\bibnamefont{Appel}} \bibnamefont{et~al.}
  (\bibinfo{collaboration}{Borexino}) (\bibinfo{year}{2022}),
  \eprint{2204.07029}.

\bibitem[{\citenamefont{{Ranucci}}(2006)}]{2006:Kamiokande}
\bibinfo{author}{\bibfnamefont{G.}~\bibnamefont{{Ranucci}}},
  \bibinfo{journal}{\prd} \textbf{\bibinfo{volume}{73}}, \bibinfo{eid}{103003}
  (\bibinfo{year}{2006}), \eprint{hep-ph/0511026}.

\bibitem[{\citenamefont{Abe et~al.}(2016)}]{Super-Kamiokande:2016yck}
\bibinfo{author}{\bibfnamefont{K.}~\bibnamefont{Abe}} \bibnamefont{et~al.}
  (\bibinfo{collaboration}{Super-Kamiokande}), \bibinfo{journal}{Phys. Rev. D}
  \textbf{\bibinfo{volume}{94}}, \bibinfo{pages}{052010}
  (\bibinfo{year}{2016}), \eprint{1606.07538}.

\bibitem[{\citenamefont{Zhuang et~al.}(2023)\citenamefont{Zhuang, Strigari,
  Jin, and Sinha}}]{Zhuang:2023dzd}
\bibinfo{author}{\bibfnamefont{Y.}~\bibnamefont{Zhuang}},
  \bibinfo{author}{\bibfnamefont{L.~E.} \bibnamefont{Strigari}},
  \bibinfo{author}{\bibfnamefont{L.}~\bibnamefont{Jin}}, \bibnamefont{and}
  \bibinfo{author}{\bibfnamefont{S.}~\bibnamefont{Sinha}}
  (\bibinfo{year}{2023}), \eprint{2307.13792}.

\bibitem[{\citenamefont{Haxton et~al.}(2013)\citenamefont{Haxton,
  Hamish~Robertson, and Serenelli}}]{2013Haxton}
\bibinfo{author}{\bibfnamefont{W.~C.} \bibnamefont{Haxton}},
  \bibinfo{author}{\bibfnamefont{R.~G.} \bibnamefont{Hamish~Robertson}},
  \bibnamefont{and} \bibinfo{author}{\bibfnamefont{A.~M.}
  \bibnamefont{Serenelli}}, \bibinfo{journal}{Ann. Rev. Astron. Astrophys.}
  \textbf{\bibinfo{volume}{51}}, \bibinfo{pages}{21} (\bibinfo{year}{2013}),
  \eprint{1208.5723}.

\bibitem[{\citenamefont{{Newstead} et~al.}(2019)\citenamefont{{Newstead},
  {Strigari}, and {Lang}}}]{2019Jayden}
\bibinfo{author}{\bibfnamefont{J.~L.} \bibnamefont{{Newstead}}},
  \bibinfo{author}{\bibfnamefont{L.~E.} \bibnamefont{{Strigari}}},
  \bibnamefont{and} \bibinfo{author}{\bibfnamefont{R.~F.}
  \bibnamefont{{Lang}}}, \bibinfo{journal}{\prd} \textbf{\bibinfo{volume}{99}},
  \bibinfo{eid}{043006} (\bibinfo{year}{2019}), \eprint{1807.07169}.

\bibitem[{\citenamefont{Suliga et~al.}(2022)\citenamefont{Suliga, Beacom, and
  Tamborra}}]{Suliga:2021hek}
\bibinfo{author}{\bibfnamefont{A.~M.} \bibnamefont{Suliga}},
  \bibinfo{author}{\bibfnamefont{J.~F.} \bibnamefont{Beacom}},
  \bibnamefont{and} \bibinfo{author}{\bibfnamefont{I.}~\bibnamefont{Tamborra}},
  \bibinfo{journal}{Phys. Rev. D} \textbf{\bibinfo{volume}{105}},
  \bibinfo{pages}{043008} (\bibinfo{year}{2022}), \eprint{2112.09168}.

\bibitem[{\citenamefont{Franco et~al.}(2016)}]{2016Franco}
\bibinfo{author}{\bibfnamefont{D.}~\bibnamefont{Franco}} \bibnamefont{et~al.},
  \bibinfo{journal}{JCAP} \textbf{\bibinfo{volume}{08}}, \bibinfo{pages}{017}
  (\bibinfo{year}{2016}), \eprint{1510.04196}.

\bibitem[{\citenamefont{{Szydagis} et~al.}(2011)\citenamefont{{Szydagis},
  {Barry}, {Kazkaz}, {Mock}, {Stolp}, {Sweany}, {Tripathi}, {Uvarov}, {Walsh},
  and {Woods}}}]{2011NEST}
\bibinfo{author}{\bibfnamefont{M.}~\bibnamefont{{Szydagis}}},
  \bibinfo{author}{\bibfnamefont{N.}~\bibnamefont{{Barry}}},
  \bibinfo{author}{\bibfnamefont{K.}~\bibnamefont{{Kazkaz}}},
  \bibinfo{author}{\bibfnamefont{J.}~\bibnamefont{{Mock}}},
  \bibinfo{author}{\bibfnamefont{D.}~\bibnamefont{{Stolp}}},
  \bibinfo{author}{\bibfnamefont{M.}~\bibnamefont{{Sweany}}},
  \bibinfo{author}{\bibfnamefont{M.}~\bibnamefont{{Tripathi}}},
  \bibinfo{author}{\bibfnamefont{S.}~\bibnamefont{{Uvarov}}},
  \bibinfo{author}{\bibfnamefont{N.}~\bibnamefont{{Walsh}}}, \bibnamefont{and}
  \bibinfo{author}{\bibfnamefont{M.}~\bibnamefont{{Woods}}},
  \bibinfo{journal}{arXiv e-prints} \bibinfo{eid}{arXiv:1106.1613}
  (\bibinfo{year}{2011}), \eprint{1106.1613}.

\bibitem[{\citenamefont{{Newstead} et~al.}(2021)\citenamefont{{Newstead},
  {Lang}, and {Strigari}}}]{2021Jayden}
\bibinfo{author}{\bibfnamefont{J.~L.} \bibnamefont{{Newstead}}},
  \bibinfo{author}{\bibfnamefont{R.~F.} \bibnamefont{{Lang}}},
  \bibnamefont{and} \bibinfo{author}{\bibfnamefont{L.~E.}
  \bibnamefont{{Strigari}}}, \bibinfo{journal}{\prd}
  \textbf{\bibinfo{volume}{104}}, \bibinfo{eid}{115022} (\bibinfo{year}{2021}),
  \eprint{2002.08566}.

\bibitem[{\citenamefont{Aprile et~al.}(2020)}]{XENON:2020rca}
\bibinfo{author}{\bibfnamefont{E.}~\bibnamefont{Aprile}} \bibnamefont{et~al.}
  (\bibinfo{collaboration}{XENON}), \bibinfo{journal}{Phys. Rev. D}
  \textbf{\bibinfo{volume}{102}}, \bibinfo{pages}{072004}
  (\bibinfo{year}{2020}), \eprint{2006.09721}.

\bibitem[{\citenamefont{{Agnes} et~al.}(2016)\citenamefont{{Agnes}, {Agostino},
  {Albuquerque}, {Alexander}, {Alton}, {Arisaka}, {Back}, {Baldin}, {Biery},
  {Bonfini} et~al.}}]{2016PhRvD..93h1101A}
\bibinfo{author}{\bibfnamefont{P.}~\bibnamefont{{Agnes}}},
  \bibinfo{author}{\bibfnamefont{L.}~\bibnamefont{{Agostino}}},
  \bibinfo{author}{\bibfnamefont{I.~F.~M.} \bibnamefont{{Albuquerque}}},
  \bibinfo{author}{\bibfnamefont{T.}~\bibnamefont{{Alexander}}},
  \bibinfo{author}{\bibfnamefont{A.~K.} \bibnamefont{{Alton}}},
  \bibinfo{author}{\bibfnamefont{K.}~\bibnamefont{{Arisaka}}},
  \bibinfo{author}{\bibfnamefont{H.~O.} \bibnamefont{{Back}}},
  \bibinfo{author}{\bibfnamefont{B.}~\bibnamefont{{Baldin}}},
  \bibinfo{author}{\bibfnamefont{K.}~\bibnamefont{{Biery}}},
  \bibinfo{author}{\bibfnamefont{G.}~\bibnamefont{{Bonfini}}},
  \bibnamefont{et~al.}, \bibinfo{journal}{\prd} \textbf{\bibinfo{volume}{93}},
  \bibinfo{eid}{081101} (\bibinfo{year}{2016}), \eprint{1510.00702}.

\bibitem[{\citenamefont{Sturrock et~al.}(2009)\citenamefont{Sturrock, Walther,
  and Wheatland}}]{Sturrock:Homestake}
\bibinfo{author}{\bibfnamefont{P.}~\bibnamefont{Sturrock}},
  \bibinfo{author}{\bibfnamefont{G.}~\bibnamefont{Walther}}, \bibnamefont{and}
  \bibinfo{author}{\bibfnamefont{S.}~\bibnamefont{Wheatland}},
  \bibinfo{journal}{The Astrophysical Journal} \textbf{\bibinfo{volume}{491}},
  \bibinfo{pages}{409} (\bibinfo{year}{2009}).

\bibitem[{\citenamefont{{Aharmim} et~al.}(2005)\citenamefont{{Aharmim},
  {Ahmed}, {Anthony}, {Beier}, {Bellerive}, {Bergevin}, {Biller}, {Boulay},
  {Bowler}, {Chan} et~al.}}]{2005:Sudbury}
\bibinfo{author}{\bibfnamefont{B.}~\bibnamefont{{Aharmim}}},
  \bibinfo{author}{\bibfnamefont{S.~N.} \bibnamefont{{Ahmed}}},
  \bibinfo{author}{\bibfnamefont{A.~E.} \bibnamefont{{Anthony}}},
  \bibinfo{author}{\bibfnamefont{E.~W.} \bibnamefont{{Beier}}},
  \bibinfo{author}{\bibfnamefont{A.}~\bibnamefont{{Bellerive}}},
  \bibinfo{author}{\bibfnamefont{M.}~\bibnamefont{{Bergevin}}},
  \bibinfo{author}{\bibfnamefont{S.~D.} \bibnamefont{{Biller}}},
  \bibinfo{author}{\bibfnamefont{M.~G.} \bibnamefont{{Boulay}}},
  \bibinfo{author}{\bibfnamefont{M.~G.} \bibnamefont{{Bowler}}},
  \bibinfo{author}{\bibfnamefont{Y.~D.} \bibnamefont{{Chan}}},
  \bibnamefont{et~al.}, \bibinfo{journal}{\prd} \textbf{\bibinfo{volume}{72}},
  \bibinfo{eid}{052010} (\bibinfo{year}{2005}), \eprint{hep-ex/0507079}.

\bibitem[{\citenamefont{{Ranucci} and {Rovere}}(2007)}]{2007:SNO}
\bibinfo{author}{\bibfnamefont{G.}~\bibnamefont{{Ranucci}}} \bibnamefont{and}
  \bibinfo{author}{\bibfnamefont{M.}~\bibnamefont{{Rovere}}},
  \bibinfo{journal}{\prd} \textbf{\bibinfo{volume}{75}}, \bibinfo{eid}{013010}
  (\bibinfo{year}{2007}), \eprint{hep-ph/0605212}.

\bibitem[{\citenamefont{Bellini et~al.}(2012)}]{Borexino:2011bhn}
\bibinfo{author}{\bibfnamefont{G.}~\bibnamefont{Bellini}} \bibnamefont{et~al.}
  (\bibinfo{collaboration}{Borexino}), \bibinfo{journal}{Phys. Lett. B}
  \textbf{\bibinfo{volume}{707}}, \bibinfo{pages}{22} (\bibinfo{year}{2012}),
  \eprint{1104.2150}.

\bibitem[{\citenamefont{{Agostini} et~al.}(2017)\citenamefont{{Agostini},
  {Altenm{\"u}ller}, {Appel}, {Atroshchenko}, {Basilico}, {Bellini},
  {Benziger}, {Bick}, {Bonfini}, {Borodikhina} et~al.}}]{2017:Borexino}
\bibinfo{author}{\bibfnamefont{M.}~\bibnamefont{{Agostini}}},
  \bibinfo{author}{\bibfnamefont{K.}~\bibnamefont{{Altenm{\"u}ller}}},
  \bibinfo{author}{\bibfnamefont{S.}~\bibnamefont{{Appel}}},
  \bibinfo{author}{\bibfnamefont{V.}~\bibnamefont{{Atroshchenko}}},
  \bibinfo{author}{\bibfnamefont{D.}~\bibnamefont{{Basilico}}},
  \bibinfo{author}{\bibfnamefont{G.}~\bibnamefont{{Bellini}}},
  \bibinfo{author}{\bibfnamefont{J.}~\bibnamefont{{Benziger}}},
  \bibinfo{author}{\bibfnamefont{D.}~\bibnamefont{{Bick}}},
  \bibinfo{author}{\bibfnamefont{G.}~\bibnamefont{{Bonfini}}},
  \bibinfo{author}{\bibfnamefont{L.}~\bibnamefont{{Borodikhina}}},
  \bibnamefont{et~al.}, \bibinfo{journal}{Astroparticle Physics}
  \textbf{\bibinfo{volume}{92}}, \bibinfo{pages}{21} (\bibinfo{year}{2017}),
  \eprint{1701.07970}.

\bibitem[{\citenamefont{{Yoo} et~al.}(2003)\citenamefont{{Yoo}, {Ashie},
  {Fukuda}, {Fukuda}, {Ishihara}, {Itow}, {Koshio}, {Minamino}, {Miura},
  {Moriyama} et~al.}}]{2003PhRvD..68i2002Y}
\bibinfo{author}{\bibfnamefont{J.}~\bibnamefont{{Yoo}}},
  \bibinfo{author}{\bibfnamefont{Y.}~\bibnamefont{{Ashie}}},
  \bibinfo{author}{\bibfnamefont{S.}~\bibnamefont{{Fukuda}}},
  \bibinfo{author}{\bibfnamefont{Y.}~\bibnamefont{{Fukuda}}},
  \bibinfo{author}{\bibfnamefont{K.}~\bibnamefont{{Ishihara}}},
  \bibinfo{author}{\bibfnamefont{Y.}~\bibnamefont{{Itow}}},
  \bibinfo{author}{\bibfnamefont{Y.}~\bibnamefont{{Koshio}}},
  \bibinfo{author}{\bibfnamefont{A.}~\bibnamefont{{Minamino}}},
  \bibinfo{author}{\bibfnamefont{M.}~\bibnamefont{{Miura}}},
  \bibinfo{author}{\bibfnamefont{S.}~\bibnamefont{{Moriyama}}},
  \bibnamefont{et~al.}, \bibinfo{journal}{\prd} \textbf{\bibinfo{volume}{68}},
  \bibinfo{eid}{092002} (\bibinfo{year}{2003}), \eprint{hep-ex/0307070}.

\bibitem[{\citenamefont{{Sturrock}}(2003)}]{2003Sturrock}
\bibinfo{author}{\bibfnamefont{P.~A.} \bibnamefont{{Sturrock}}},
  \bibinfo{journal}{\apj} \textbf{\bibinfo{volume}{594}}, \bibinfo{pages}{1102}
  (\bibinfo{year}{2003}), \eprint{hep-ph/0304073}.

\bibitem[{\citenamefont{{Sturrock} et~al.}(2005)\citenamefont{{Sturrock},
  {Caldwell}, {Scargle}, and {Wheatland}}}]{2005Sturrock}
\bibinfo{author}{\bibfnamefont{P.~A.} \bibnamefont{{Sturrock}}},
  \bibinfo{author}{\bibfnamefont{D.~O.} \bibnamefont{{Caldwell}}},
  \bibinfo{author}{\bibfnamefont{J.~D.} \bibnamefont{{Scargle}}},
  \bibnamefont{and} \bibinfo{author}{\bibfnamefont{M.~S.}
  \bibnamefont{{Wheatland}}}, \bibinfo{journal}{arXiv e-prints}
  \bibinfo{eid}{hep-ph/0501205} (\bibinfo{year}{2005}),
  \eprint{hep-ph/0501205}.

\bibitem[{\citenamefont{{Hosaka} et~al.}(2006)\citenamefont{{Hosaka},
  {Ishihara}, {Kameda}, {Koshio}, {Minamino}, {Mitsuda}, {Miura}, {Moriyama},
  {Nakahata}, {Namba} et~al.}}]{2006:SK1}
\bibinfo{author}{\bibfnamefont{J.}~\bibnamefont{{Hosaka}}},
  \bibinfo{author}{\bibfnamefont{K.}~\bibnamefont{{Ishihara}}},
  \bibinfo{author}{\bibfnamefont{J.}~\bibnamefont{{Kameda}}},
  \bibinfo{author}{\bibfnamefont{Y.}~\bibnamefont{{Koshio}}},
  \bibinfo{author}{\bibfnamefont{A.}~\bibnamefont{{Minamino}}},
  \bibinfo{author}{\bibfnamefont{C.}~\bibnamefont{{Mitsuda}}},
  \bibinfo{author}{\bibfnamefont{M.}~\bibnamefont{{Miura}}},
  \bibinfo{author}{\bibfnamefont{S.}~\bibnamefont{{Moriyama}}},
  \bibinfo{author}{\bibfnamefont{M.}~\bibnamefont{{Nakahata}}},
  \bibinfo{author}{\bibfnamefont{T.}~\bibnamefont{{Namba}}},
  \bibnamefont{et~al.}, \bibinfo{journal}{\prd} \textbf{\bibinfo{volume}{73}},
  \bibinfo{eid}{112001} (\bibinfo{year}{2006}), \eprint{hep-ex/0508053}.

\bibitem[{\citenamefont{{Abe} et~al.}(2023)\citenamefont{{Abe}, {Bronner},
  {Hayato}, {Hiraide}, {Hosokawa}, {Ieki}, {Ikeda}, {Kameda}, {Kanemura},
  {Kaneshima} et~al.}}]{Kamioka:2023}
\bibinfo{author}{\bibfnamefont{K.}~\bibnamefont{{Abe}}},
  \bibinfo{author}{\bibfnamefont{C.}~\bibnamefont{{Bronner}}},
  \bibinfo{author}{\bibfnamefont{Y.}~\bibnamefont{{Hayato}}},
  \bibinfo{author}{\bibfnamefont{K.}~\bibnamefont{{Hiraide}}},
  \bibinfo{author}{\bibfnamefont{K.}~\bibnamefont{{Hosokawa}}},
  \bibinfo{author}{\bibfnamefont{K.}~\bibnamefont{{Ieki}}},
  \bibinfo{author}{\bibfnamefont{M.}~\bibnamefont{{Ikeda}}},
  \bibinfo{author}{\bibfnamefont{J.}~\bibnamefont{{Kameda}}},
  \bibinfo{author}{\bibfnamefont{Y.}~\bibnamefont{{Kanemura}}},
  \bibinfo{author}{\bibfnamefont{R.}~\bibnamefont{{Kaneshima}}},
  \bibnamefont{et~al.}, \bibinfo{journal}{arXiv e-prints}
  \bibinfo{eid}{arXiv:2311.01159} (\bibinfo{year}{2023}), \eprint{2311.01159}.

\bibitem[{\citenamefont{{Heix} et~al.}(2019)\citenamefont{{Heix}, {Tilav},
  {Wiebusch}, and {Z{\"o}cklein}}}]{2019IceCube}
\bibinfo{author}{\bibfnamefont{P.}~\bibnamefont{{Heix}}},
  \bibinfo{author}{\bibfnamefont{S.}~\bibnamefont{{Tilav}}},
  \bibinfo{author}{\bibfnamefont{C.}~\bibnamefont{{Wiebusch}}},
  \bibnamefont{and}
  \bibinfo{author}{\bibfnamefont{M.}~\bibnamefont{{Z{\"o}cklein}}},
  \bibinfo{journal}{arXiv e-prints} \bibinfo{eid}{arXiv:1909.02036}
  (\bibinfo{year}{2019}), \eprint{1909.02036}.

\bibitem[{\citenamefont{{Borexino Collaboration}
  et~al.}(2013)\citenamefont{{Borexino Collaboration}, {Bellini}, {Benziger},
  {Bick}, {Bonfini}, {Bravo}, {Avanzini}, {Caccianiga}, {Cadonati}, {Calaprice}
  et~al.}}]{2013Borexino}
\bibinfo{author}{\bibnamefont{{Borexino Collaboration}}},
  \bibinfo{author}{\bibfnamefont{G.}~\bibnamefont{{Bellini}}},
  \bibinfo{author}{\bibfnamefont{J.}~\bibnamefont{{Benziger}}},
  \bibinfo{author}{\bibfnamefont{D.}~\bibnamefont{{Bick}}},
  \bibinfo{author}{\bibfnamefont{G.}~\bibnamefont{{Bonfini}}},
  \bibinfo{author}{\bibfnamefont{D.}~\bibnamefont{{Bravo}}},
  \bibinfo{author}{\bibfnamefont{M.~B.} \bibnamefont{{Avanzini}}},
  \bibinfo{author}{\bibfnamefont{B.}~\bibnamefont{{Caccianiga}}},
  \bibinfo{author}{\bibfnamefont{L.}~\bibnamefont{{Cadonati}}},
  \bibinfo{author}{\bibfnamefont{F.}~\bibnamefont{{Calaprice}}},
  \bibnamefont{et~al.}, \bibinfo{journal}{arXiv e-prints}
  \bibinfo{eid}{arXiv:1308.0443} (\bibinfo{year}{2013}), \eprint{1308.0443}.

\bibitem[{\citenamefont{{Scargle}}(1982)}]{1982Scargle}
\bibinfo{author}{\bibfnamefont{J.~D.} \bibnamefont{{Scargle}}},
  \bibinfo{journal}{\apj} \textbf{\bibinfo{volume}{263}}, \bibinfo{pages}{835}
  (\bibinfo{year}{1982}).

\bibitem[{\citenamefont{VanderPlas}(2018)}]{2018VanderPlas}
\bibinfo{author}{\bibfnamefont{J.~T.} \bibnamefont{VanderPlas}},
  \bibinfo{journal}{The Astrophysical Journal Supplement Series}
  \textbf{\bibinfo{volume}{236}}, \bibinfo{pages}{16} (\bibinfo{year}{2018}).

\bibitem[{\citenamefont{Virtanen et~al.}(2020)\citenamefont{Virtanen, Gommers,
  Oliphant, Haberland, Reddy, Cournapeau, Burovski, Peterson, Weckesser, Bright
  et~al.}}]{2020SciPy-NMeth}
\bibinfo{author}{\bibfnamefont{P.}~\bibnamefont{Virtanen}},
  \bibinfo{author}{\bibfnamefont{R.}~\bibnamefont{Gommers}},
  \bibinfo{author}{\bibfnamefont{T.~E.} \bibnamefont{Oliphant}},
  \bibinfo{author}{\bibfnamefont{M.}~\bibnamefont{Haberland}},
  \bibinfo{author}{\bibfnamefont{T.}~\bibnamefont{Reddy}},
  \bibinfo{author}{\bibfnamefont{D.}~\bibnamefont{Cournapeau}},
  \bibinfo{author}{\bibfnamefont{E.}~\bibnamefont{Burovski}},
  \bibinfo{author}{\bibfnamefont{P.}~\bibnamefont{Peterson}},
  \bibinfo{author}{\bibfnamefont{W.}~\bibnamefont{Weckesser}},
  \bibinfo{author}{\bibfnamefont{J.}~\bibnamefont{Bright}},
  \bibnamefont{et~al.}, \bibinfo{journal}{Nature Methods}
  \textbf{\bibinfo{volume}{17}}, \bibinfo{pages}{261} (\bibinfo{year}{2020}).

\bibitem[{\citenamefont{Price-Whelan et~al.}(2018)\citenamefont{Price-Whelan,
  Sip{\H{o}}cz, G{\"u}nther, Lim, Crawford, Conseil, Shupe, Craig, Dencheva,
  Ginsburg et~al.}}]{astropy:2013}
\bibinfo{author}{\bibfnamefont{A.~M.} \bibnamefont{Price-Whelan}},
  \bibinfo{author}{\bibfnamefont{B.}~\bibnamefont{Sip{\H{o}}cz}},
  \bibinfo{author}{\bibfnamefont{H.}~\bibnamefont{G{\"u}nther}},
  \bibinfo{author}{\bibfnamefont{P.}~\bibnamefont{Lim}},
  \bibinfo{author}{\bibfnamefont{S.}~\bibnamefont{Crawford}},
  \bibinfo{author}{\bibfnamefont{S.}~\bibnamefont{Conseil}},
  \bibinfo{author}{\bibfnamefont{D.}~\bibnamefont{Shupe}},
  \bibinfo{author}{\bibfnamefont{M.}~\bibnamefont{Craig}},
  \bibinfo{author}{\bibfnamefont{N.}~\bibnamefont{Dencheva}},
  \bibinfo{author}{\bibfnamefont{A.}~\bibnamefont{Ginsburg}},
  \bibnamefont{et~al.}, \bibinfo{journal}{The Astronomical Journal}
  \textbf{\bibinfo{volume}{156}}, \bibinfo{pages}{123} (\bibinfo{year}{2018}).

\bibitem[{\citenamefont{Mishra and Strigari}(2023)}]{Mishra:2023jlq}
\bibinfo{author}{\bibfnamefont{N.}~\bibnamefont{Mishra}} \bibnamefont{and}
  \bibinfo{author}{\bibfnamefont{L.~E.} \bibnamefont{Strigari}},
  \bibinfo{journal}{Phys. Rev. D} \textbf{\bibinfo{volume}{108}},
  \bibinfo{pages}{063023} (\bibinfo{year}{2023}), \eprint{2305.17827}.

\end{thebibliography}

\end{document}